\documentclass[prd,aps,floats,floatfix,superscriptaddress,preprintnumbers,eqsecnum,nofootinbib,twocolumn]{revtex4}
\usepackage{amsfonts,amsmath,epsfig,amssymb,latexsym,color,graphicx,array,subfigure,bm,float }

\definecolor{red}{rgb}{1,0,0}
\definecolor{blue }{rgb}{0,0,1}
\definecolor{green}{rgb}{0,1,0}


\newcommand{\bea}{\begin{eqnarray}}
\newcommand{\ena}{\end{eqnarray}}
\newcommand{\beann}{\begin{eqnarray*}}
\newcommand{\enann}{\end{eqnarray*}}

\newcommand{\dsl}{\pa \kern-0.5em /}

\newcommand{\pa}{\partial}

\newcommand{\vect}[1]{\!\!\!\mbox{ \boldmath $#1$}}

\newcommand{\gYM}{g_{\scriptscriptstyle{\rm YM}}}
\newcommand{\rhoYM}{\rho_{\scriptscriptstyle{\rm YM}}}
\newcommand{\OmYM}{\Omega_{\scriptscriptstyle{\rm YM}}}
\newcommand{\EYM}{E_{\scriptscriptstyle{\rm YM}}}
\newcommand{\BYM}{B_{\scriptscriptstyle{\rm YM}}}
\newcommand{\SYM}{S_{\scriptscriptstyle{\rm YM}}}

\begin{document}


\title{Inflationary Dynamics with a Non-Abelian Gauge Field}

\author{Kei-ichi Maeda}
\email{maeda"at"waseda.jp}
\affiliation{Department of Applied Mathematics and Theoretical Physics, 
University of Cambridge,
Wilberforce Road, Cambridge CB3 0WA, United Kingdom}
\affiliation{APC-AstroParticule et Cosmologie {\rm (CNRS-Universit$\acute{e}$ Paris 7)}, 
10 rue Alice Domon et L$\acute{e}$onie Duquet, 75205 Paris Cedex 13, France}
\affiliation{
Department of Physics, Waseda University,
Shinjuku, Tokyo 169-8555, Japan}

\author{Kei Yamamoto}
\email{K.Yamamoto"at"damtp.cam.ac.uk}
\affiliation{Department of Applied Mathematics and Theoretical Physics, 
University of Cambridge,
Wilberforce Road, Cambridge CB3 0WA, United Kingdom}
\affiliation{Institute of Theoretical Astrophysics, University of Oslo, \\
P.O. Box 1029, Blindern, N-0315 Oslo, Norway}

\begin{abstract}
We study the dynamics of the universe with a scalar field and an SU(2) 
non-Abelian Gauge (Yang-Mills) field. 
The scalar field has an exponential potential and the Yang-Mills field
is coupled to the scalar field with an exponential function of the scalar 
field. 
We find that the magnetic component of the Yang-Mills
 field assists  acceleration of the cosmic expansion
 and a power-law inflation becomes possible even if 
the scalar field potential is steep, which may be expected from 
some compactification of higher-dimensional unified theories of 
fundamental interactions.
This power-law inflationary solution is a stable attractor
in a certain range of coupling parameters.
Unlike the case with multiple Abelian gauge
 fields, the power-law inflationary solution
with the dominant electric component is unstable because of 
the existence of non-linear coupling of the Yang-Mills field.
We also analyze the dynamics for the 
non-inflationary regime, 
and find several attractor solutions.
\end{abstract}

\maketitle




\section{Introduction}
\label{introduction}
The idea of inflation  now gives a standard scenario of  
 the early evolution of the universe
\cite{Starobinsky,inflation1,inflation2,inflation3,inflation4}. 
It solves several difficulties such as the horizon and 
flatness problems in the Big-Bang cosmology, which has been 
confirmed by the precision cosmological observations.
It also provides us  with a prediction on the origin 
of the observed density fluctuations.
Many cosmological models with such a phase of 
accelerated expansion have been proposed by 
introducing a scalar field with an appropriate potential
(or some alternative fields). 
However, it is desirable to derive a natural
model from a fundamental theory of particle
physics without introducing any   anonymous field
by hand.
The most promising candidate
for such a fundamental theory
is the ten-dimensional superstring theory\cite{string} 
or eleven-dimensional M-theory\cite{Mtheory}.
They are hoped to give an interesting explanation
for the accelerated expansion of the 
universe upon compactification to four dimensions.

In the low-energy effective field theories of superstrings 
or supergravity theories, however, 
there is the so-called no-go theorem, which forbids such
 an inflating solution 
if the internal space is a time-independent 
nonsingular compact manifold without boundary~\cite{no-go}. 
In order to  evade this theorem,
we have to violate some of those assumptions.
We have three possibilities:\\

\begin{itemize}
\item a time-dependent internal space such as 
S-branes\cite{Sbrane1,Sbrane2,Sbrane3,Sbrane4} 
\item an introduction of ``singularity" such as branes\cite{Brane1,Brane2} 
\item a modification of gravitational action 
such as higher-curvature 
terms\cite{Starobinsky,HighR1,HighR2,HighR3,HighR4,HighR5} 
\end{itemize}

Although some models could be promising, many models are still suffering 
from  instability of a dilaton field or moduli fields.
In fact, we naturally expect exponential couplings of moduli fields.
Without fixing those moduli, many inflationary models are spoiled.

An exponential coupling is not always harmful for inflation, however.
In fact, we can find a power-law inflation\cite{power-law_inflation}
 with an exponential potential\cite{HAL,Yokoyama_Maeda}. 
 It also provides the cosmic no hair theorem
similar to the slow-roll inflation\cite{Kitada_Maeda}.
In supergravity theories and superstring models, 
an effective exponential potential $V_0 \exp[-\alpha\phi]$ 
 naturally appears\cite{SG1, Maeda_Nishino, SG2}. 
However, their potential
is usually so steep that the power exponent of the scale
factor cannot be much larger than unity, which makes
it difficult to construct an acceptable inflationary
model of the universe. For example, we find $\alpha=\sqrt{2}$ and
$\sqrt{6}$ for two scalar fields in $N= 2$, six-dimensional supergravity
model with $S^2$-compactification\cite{Maeda_Nishino}, and
the same is true for two scalar fields in $N= 1$, ten-dimensional
supergravity model with gaugino condensation
\cite{SG2}. 
Townsend summarized the possible exponential potentials 
derived by the compactification of 
ten- or eleven-dimensional supergravity theories\cite{Townsend}. 
From flux compactifications, one expects $\alpha\geq \sqrt{6}$,
while we may find $\sqrt{2}\leq \alpha \leq \sqrt{6}$ 
by hyperbolic compactifications.
Neither of them offers a flat enough potential for inflation.

In the unified theories of fundamental interactions, there  
naturally exist gauge fields, which may be included in the original
action such as the heterotic string theory or can be induced by 
Kaluza-Klein compactification.
In effective four-dimensional theories derived from higher-dimensional
unified theories, we also expect those gauge fields coupled exponentially 
to moduli fields such as ${1\over 4}\exp[\lambda\phi]\,\vect{F}^2$.
Hull and Townsend discussed such a coupling 
for the case of U(1) gauge fields. They found that 
the possible values of the coupling in the four-dimensional 
effective action are $\lambda=0, 
\sqrt{2/3}, \sqrt{2}$, or $\sqrt{6}$ in the context of 
black holes in the type II string theory compactified 
on a six torus\cite{Hull-Townsend}.
In M-theory (eleven-dimensional supergravity) with intersecting branes, 
the four-dimensional effective action 
also contains the same moduli couplings to U(1) multiplet
\cite{Gibbons-Maeda}.

 If the strengths of the couplings between gauge fields and a scalar
field are similar to that of the scalar self-coupling, the gauge fields may
affect the dynamics of the scalar field.
In fact there are several discussions about the dynamics of inflation, 
 where supportive roles of gauge fields in realizing accelerated 
expansion have been observed\cite{Kanno1,Watanabe,Kanno2,
Yamamoto,Murata,Moniz,Do,Hassan,Wagstaff,Hervik,Adshead,Anber}.

The effect of the gauge-kinetic coupling on the inflationary dynamics 
was first discussed 
in the context of anisotropic inflation\cite{Kanno1},
assuming a U(1) gauge field coupled to an inflaton field.
Since a single U(1) field cannot exist in Friedmann-Lema\^itre-Robertson-Walker
(FLRW) isotropic and homogeneous spacetimes, 
they discussed Bianchi spacetimes as the cosmological model. 
 They specified the scalar potential to be quadratic and chose
$\exp[c\phi^2]$ as the gauge kinetic coupling. They showed 
that an anisotropic inflationary era may  arise as a transient 
attractor state while the scalar inflaton is slowly rolling.
The anisotropy eventually disappears as the scalar field 
oscillates towards the end of inflation.
The observational relic of the anisotropic inflationary era 
 was also discussed\cite{Kanno1,Watanabe}.

While the chaotic inflation driven by the quadratic potential
is phenomenologically interesting as it automatically results
in reheating, the form of the inflaton-gauge interaction discussed
in \cite{Kanno1} may not naturally appear in the unified theories.
They also studied the case with an exponential potential 
and a U(1) gauge field coupled exponentially to the scalar field,
 which suits the framework of the unified theories better.
They found an exact anisotropic inflationary solution, which is 
an attractor  independent of the initial conditions\cite{Kanno2}. 
Since our present universe is almost isotropic,
this model must be  severely constrained.

However, if there exist more than two gauge fields, 
we find an interesting scenario. 
Although   it requires an artificial assumption that all the
gauge fields couple to the inflaton through a common
gauge-kinetic function, one can obtain a totally homogeneous 
and isotropic inflationary solution as an attractor\cite{Yamamoto}.
Since the anisotropic inflation can be found as a transient 
attractor, we might have a chance to find distinct observational
signatures. An important result is that 
an isotropic power-law inflationary solution appears as an attractor 
even for a steep exponential potential for the inflaton, which is 
expected from the unified theories of fundamental interactions.
  While there are certain conditions to be satisfied by the 
 gauge-kinetic coupling constant, they are not so strict as the
 usual slow-roll conditions and could fall within the reach of
 the supergravity theories.

In the case of U(1) multiplet fields, we usually expect 
the different gauge-kinetic coupling constants for different
fields in the context of the unified theories.
However, if we consider a non-Abelian gauge field,
it consists of ``multiple" vector fields with a single common
gauge-kinetic coupling constant.
As a result, the discouraging feature of U(1) multiplet will disappear.
The conventional chaotic inflationary model with a non-Abelian gauge 
field has been studied\cite{Murata}.
 Motivated by its phenomenological development and the 
aforementioned features of high-energy physics, in this paper, 
we study SU(2) non-Abelian gauge field coupled 
exponentially to a scalar field with an exponential potential,
in order to know whether the non-Abelian gauge field has the similar 
nice properties as the U(1) multiplet case.

We should note that there is also an approach different from 
the present gauge-kinetic coupling model\cite{Adshead,Anber}.
They consider an axion field coupled to a non-Abelian gauge field,
which is named chromo-natural inflation.
It may give another interesting inflationary regime
with non-Abelian gauge fields.

In the following, we present the basic equations of our system,
and obtain power-law solutions in \S. \ref{solutions}.
We find that a power-law inflationary solution is found only for 
the case of the magnetic component dominance in contrast to
the U(1) triplet case, in which both inflationary solutions 
with electric field and magnetic field are possible.
In \S. \ref{Stability_Analysis}, we describe
the solutions as fixed points of a dynamical system
and analyze their stabilities. 
In \S. \ref{Numerical_Study},
we perform numerical analysis for the range of 
coupling constants where fixed points do not exist.
It also tells us  how the attractor state is achieved from 
generic initial data.
Concluding remarks and discussions are given in \S. 
\ref{Concluding_Remarks}.

\begin{widetext}
\section{Basic Equations}
\label{basic_equations}
We use the unit of $\kappa^2=8\pi G=1$. 
The action we discuss is
\begin{eqnarray*}
S=\int d^4 x\sqrt{-g}\left[{1\over 2}R-{1\over 2}(\nabla\phi)^2-V(\phi)
-{1\over 4}f^2(\phi)\, F_{\mu\nu}^{\rm (a)}F^{{\rm(a)}\mu\nu}\right]
\,,
\end{eqnarray*}
where
\begin{eqnarray*}
F_{\mu\nu}^{\rm (a)}&=&\partial_\mu A_{\nu}^{\rm (a)}
-\partial_\nu A_{\mu}^{\rm (a)}+ \gYM \epsilon_{\rm abc}
A_{\mu}^{\rm (b)}A_{\nu}^{\rm (c)}
\end{eqnarray*}
is an SU(2) non-Abelian gauge field, which we call the 
Yang-Mills (YM) field, and $\gYM$ 
is its coupling constant.
The coupling to the scalar field $f(\phi )$ and the scalar potential $V(\phi )$ 
are given respectively by
\end{widetext}
\begin{eqnarray*}
f^2(\phi)&=& e^{\lambda\phi}
\,,
\\
V(\phi)&=&V_0 e^{-\alpha \phi}
\,.
\end{eqnarray*}
$\alpha $ can be set non-negative without loss of generality.
We also restrict ourselves to $V_0 \geq 0$ since our primary 
interest here is inflation.

Throughout the article, we discuss a flat FLRW spacetime\cite{footnote1}, whose metric is
given by
\begin{eqnarray*}
ds^2=-dt^2+a^2(t)d^2\,\vect{x}
\,.
\end{eqnarray*}
We assume that the vector potential is given by
\begin{eqnarray}
A_{0}^{\rm (a)}=0\,,A_{i}^{\rm (a)}=A(t)\delta_{i}^{\rm (a)}
\,,
\label{isotropic_vector}
\end{eqnarray}
so that the  YM field is taken to be isotropic.
This configulation 
results in both homogeneous electric and magnetic components,
which are written in the coordinate basis as
\begin{eqnarray*}
E_{i}^{\rm (a)}&:=&F_{i0}^{\rm (a)} = a (t) E(t)\delta_{i}^{\rm (a)}\,,
\nonumber \\
B^{{\rm (a)}\,i}&:=&{1\over 2}  \epsilon^{ijk}F_{jk}^{\rm (a)}  
={B(t) \over a(t)}\delta^{i{\rm (a)}} , 
\end{eqnarray*}
with 
\begin{equation*}
E:=-{\dot{A}\over a}\,, \ \ \ {\rm and} \ \ \  B =  \gYM \frac{A^2}{a^2} 
\end{equation*}
being their comoving field strengths. This is an important difference from
 U(1) gauge
fields, for which we find only the electric component in the above vector 
potential. 
The homogeneous magnetic component in U(1) gauge fields is obtained 
only when we introduce an
 appropriate inhomogeneous vector potential.
As a result, the electric component and magnetic one
 in the U(1) fields are independent. 
We can discuss each component separately.
In contrast, the YM field always consists of two components 
in the above isotropic configuration (\ref{isotropic_vector})
and the homogeneous field is found only by a homogeneous vector
potential. If one introduces any spatial dependence to the vector potential, 
the field strengths become inhomogeneous.
We should also note that we need more than two U(1)  
fields with  a common  coupling to the scalar field as discussed 
in \cite{Yamamoto}
in order to find an isotropic and homogeneous attractor spacetime.
Otherwise, we find an anisotropic universe.  For an SU(2) gauge field, 
this uniform coupling is a necessary consequence of the symmetry.

The Einstein equations are
\bea
&&
H^2={1\over 3}\left[
{1\over 2}\dot\phi^2+V+\rhoYM
\right]
\,,
\label{Friedmann}
\\
&&
\dot H=-\left[{1\over 2}\dot\phi^2+
{2\over 3} \rhoYM
\right]
\label{Einstein_eq2}
\,,
\ena
where the dots denote the time derivative $d/dt$, and 
$H=\dot a/a$ is the Hubble expansion rate.
$\rhoYM$ is the YM energy density, which consists of 
 the electric and magnetic parts,
i.e., 
\bea
\rhoYM=\rho_E+\rho_B\, .
\ena 
They are defined by
\bea
\rho_E={3\over 2}e^{\lambda\phi}
E^2\,,~~
\rho_B={3\over 2}e^{\lambda\phi}
B^2
\,.
\ena

The equation of motion for  the scalar field is
\bea
\ddot \phi+3H\dot \phi-\alpha V-\lambda \left(\rho_E-\rho_B
\right)=0
\,,
\label{eq_scalar}
\ena
and the equation of motion for the YM field is simply 
\bea
\ddot A+H \dot A+\lambda \dot \phi \dot A
+2\gYM^2{A^3\over a^2}=0
\label{eq_Yang-Mills}
\,.
\ena
Using the Bianchi identity, Eq. (\ref{Einstein_eq2}) is obtained from 
Eqs. (\ref{Friedmann}), (\ref{eq_scalar}) and (\ref{eq_Yang-Mills}).
Hence we take (\ref{Friedmann}), (\ref{eq_scalar}) and (\ref{eq_Yang-Mills})
as the basic equations of our system.

The YM equation can be reduced to the first order equations 
for each energy density as
\bea
\dot \rho_E&=&-(4H+\lambda\dot \phi)\rho_E -4{\dot A\over A}\rho_B
\nonumber \\
\dot \rho_B&=&-(4H-\lambda\dot \phi)\rho_B +4{\dot A\over A}\rho_B
\label{eq_YM_density}
\ena
The terms with $4H$ come from the radiation-like behavior of the YM field
($\rho_{\rm rad}\propto a^{-4}$) and the last terms are from 
the non-linear interaction in the YM field.
In fact, for U(1) triplet fields with  a uniform
exponential coupling to a scalar field,
we find the evolution equations for energy densities by dropping
the non-linear interaction terms.
As we shall see later, the non-linear terms play a significant role
in the dynamics of the model.

\section{Power-Law Solutions}
\label{solutions}
Since we have the exponential potential, we expect a power-law expansion
and look for the possibility of power-law inflation.
Suppose our solution is given by 
\begin{eqnarray}
a&=&a_0 t^p
\label{power-law}
\\
\phi&=&{2\over \alpha}\ln t+\phi_0
\label{evol_scalar}
\,,
\end{eqnarray}
where $p$ is assumed to be a constant, and $a_0$ and $\phi_0$ are 
initial values.
 The coefficient $2/\alpha$ in front of $\ln t$ is determined by 
requiring that the $t$ dependence of $\dot \phi^2$ and $V$ be the same,
in order to satisfy the Hamiltonian constraint (\ref{Friedmann}).

\subsection{The case with U(1) triplet fields}
\label{power_law_U1}
First we consider the case  with U(1)  triplet fields, which was 
discussed in \cite{Yamamoto}.
The equations to be solved are
\bea
\dot \rho_E&=&-(4H+\lambda\dot \phi)\rho_E 
\nonumber \\
\dot \rho_B&=&-(4H-\lambda\dot \phi)\rho_B
\label{eq_EM}
\,,
\ena
and Eqs. (\ref{Friedmann}) and (\ref{eq_scalar}).
In our setting (\ref{isotropic_vector}), the magnetic field vanishes.
However, if we add an appropriate inhomegeneoes vector potential,
a homegeneous magnetic field can appear and the energy densities of
the electromagnetic fields 
satisfy Eqs. (\ref{eq_EM}).

\subsubsection{Dynamics of the scalar field}
Eqs. (\ref{eq_EM}) are eaily integrated as
\bea
\rho_E&=&\rho_{E0}{e^{-\lambda(\phi-\phi_0)}\over (a/a_0)^4} 
\nonumber \\
\rho_B&=&\rho_{B0}{e^{\lambda(\phi-\phi_0)}\over (a/a_0)^4} 
\label{sol_EM}
\ena
where $\rho_{E0}, \rho_{B0}, \phi_0$ and $a_0$ are integration constants.

Then the equation of the scalar field is reduced to
\bea
\ddot \phi+3H\dot \phi+{\partial V_{\rm eff} \over \partial \phi}=0
\,,
\ena
where
\bea
 V_{\rm eff} :=V_0 e^{-\alpha\phi}+{1\over a^4}
\left(C_E e^{-\lambda\phi}+C_B e^{\lambda\phi}\right)
\label{potential_eff}
\ena
with
\bea
C_E=\rho_{E0}a_0^4 e^{\lambda\phi_0}\,,~~
C_B=\rho_{B0}a_0^4 e^{-\lambda\phi_0}
\,.
\ena
Although the original potential $V$ is monotonically decreasing,
the effective potential (\ref{potential_eff}) has a minimum point
for $\lambda<0$ and $C_E\neq 0$, or 
for $\lambda>0$ and $C_B\neq 0$.
As a result, the scalar field will evolve more slowly than the case only with 
the exponential potential $V$. Since there exists the pre-factor $a^{-4}$,
the minimum point will move and the minimum value will decrease 
as the universe evolves. Hence we do not have an exponential 
expansion, but have a power-law expansion whose power exponent is
larger than the original power-law expansion driven solely by the potential $V$.
This is the mechanism that a gauge field coupled to a scalar field assists
slowing down the motion of the scalar field and 
inflationary expansion becomes possible even for 
a steep potential.

Next we present the explicit power-law solutions.
Assuming that the expansion of the universe and the evolution of 
the scalar field are described by Eqs. (\ref{power-law}) and 
(\ref{evol_scalar}), and 
the energy densities of the electromagnetic fields
are proportional to $t^{-2}$, i.e., 
$\rho_{E}=\rho_{E0}/t^2$ and $\rho_{B}=\rho_{B0}/t^2$,
we find that Eqs.(\ref{eq_EM}) become the algebraic equations:
\bea
&&
\rho_{E0}=\left(2p+{\lambda\over \alpha}\right)\rho_{E0}
\label{electric_energy_density}
\,,
\\
&&
\rho_{B0}=\left(2p-{\lambda\over \alpha}\right)\rho_{B0}
\label{magnetic_energy_density}
\,.
\ena
There are two cases:$\rho_{B0}=0$ and $\rho_{E0}=0$,
 which we shall discuss separately.

\subsubsection{The case with the electric field ${\rm (}E_{\rm U1}${\rm )}}
\label{EU1}

For the case with the dominant electric field ($\rho_B=0$),
which we shall call the regime $E_{\rm U1}$,
assuming $\rho_E=\rho_{E0}t^{-2}$ ($\rho_{E0}$: constant),
we find three algebraic equations: Eq. (\ref{electric_energy_density}) 
and 
\bea
&&
p^2={1\over 3}\left({2\over \alpha^2}+V_0e^{-\alpha\phi_0}
+\rho_{E0}\right)
\\
&&
-{2\over \alpha}+{6p\over \alpha}-\alpha V_0e^{-\alpha\phi_0}
\lambda \rho_{E0}=0
\ena
which are rearranged into
\bea
&&
p={1\over 2}\left(1-{\lambda\over \alpha}\right)
\label{U1E_p}
\\
&&
V_0e^{-\alpha\phi_0}={1\over 4\alpha^2}\left[4-3\lambda(\alpha-\lambda)\right]
\label{U1E_V}
\\
&&
\rho_{E0}={3\over 4\alpha^2}\left[\alpha(\alpha-\lambda)-4\right]
\,.
\label{U1E_E}
\ena
Since the left-hand-sides are positive definite, in order for such a solution 
to exist, we have to impose the following conditions:
\bea
\lambda
\leq
\left\{
\begin{array}{cc}
\alpha-{4\over \alpha} & (\alpha\leq \sqrt{6})\\
{1\over 2}\left(\alpha-\sqrt{\alpha^2-16/3}\right) 
& (\alpha\geq \sqrt{6})\\
\end{array}
\right.
\label{cond_U1E}
\ena

The power-law inflation is possible for the range of 
coupling parameters of
$\lambda<-\alpha$ and $\alpha(\alpha-\lambda)>4$, 
as was shown in \cite{Yamamoto}.

\subsubsection{The case with the magnetic field ${\rm (}B_{\rm U1}${\rm )}}
\label{MU1}

For the case only with the magnetic field ($\rho_{E}=0$),
which we shall call the regime $B_{\rm U1}$,
we find the same result by changing the sign of $\lambda$.
The solution is described by
\bea
&&
p={1\over 2}\left(1+{\lambda\over \alpha}\right)
\label{U1B_p}\\
&&
V_0e^{-\alpha\phi_0}={1\over 4\alpha^2}\left[4+3\lambda(\alpha+\lambda)\right]
\label{U1B_V}\\
&&
\rho_{B0}={3\over 4\alpha^2}\left[\alpha(\alpha+\lambda)-4\right]
\label{U1B_B}\,,
\ena
and the existence conditions are
\bea
\lambda
\geq 
\left\{
\begin{array}{cc}
-\alpha+{4\over \alpha} & (\alpha\leq \sqrt{6})\\
-{1\over 2}\left(\alpha-\sqrt{\alpha^2-16/3}\right) 
& (\alpha\geq \sqrt{6})\\
\end{array}
\right.
\label{cond_U1B}
\ena
The power-law inflation is obtained for the parameter
range of $\lambda>\alpha$ and $\alpha(\alpha+\lambda)>4$. 

\begin{figure}[h]
\begin{center}
\includegraphics[scale=.4]{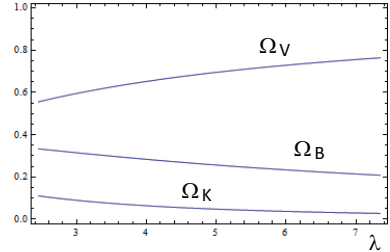}
\caption{The density parameters of each component
[$\Omega_{B}=$(the magentic field), $\Omega_V$
(the potential), and $\Omega_K$(the kinetic term of the scalar field)]
for the case of $\alpha=\sqrt{6}$ and $\lambda>\alpha$.
The power exponent of the scale factor is given by $p=(1+\lambda/\alpha)/2$.
}
\label{fig:density_parameter_U(1)}
\end{center}
\end{figure}

Defining 
the density parameters of each component by
$\Omega_{E}=\rho_E/3H^2$ (the electric field), $\Omega_{B}=\rho_B/3H^2$
(the magnetic field),
$\Omega_V=V/3H^2$ (the potential), and $\Omega_K=\dot \phi^2/6H^2$
(the kinetic term of the scalar field),
we find that those depend on the coupling parameters.
We show one example for the power-law inflation with 
magnetic field in Fig. \ref{fig:density_parameter_U(1)}.
We find that the magnetic field gives a certain contribution
to the expansion of the universe.

We will show the stability condition later in \S. \ref{U(1)}.

\subsection{The case with YM field}
\label{sol_YM}
Now we show the dynamics changes when the YM interaction
is turned on.
Since we have the non-linear coupling in the YM field, a 
simple power-law ansatz may not work.
But  we first 
look for a solution similar to those found in the U(1) triplet  case.
\subsubsection{The case with the dominant electric component
${\rm (}E_{\scriptscriptstyle{\rm YM}}${\rm )}}
\label{EYM}
If we assume $\rho_E\gg\rho_B$ but $\rho _B$ being non-vanishing 
due to the interaction between electric and magnetic fields, which 
we shall call the regime $E_{\scriptscriptstyle{\rm YM}}$,
 dropping the term with $\rho_B$, 
we find the same equations as the U(1) triplet  case. 
As a result, we find the same solution [Eqs. (\ref{U1E_p})-(\ref{U1E_E})]
 as long as the electric component
stays dominant. Under the conditions
$\lambda<-\alpha$ and $\alpha(\alpha-\lambda)>4$, 
we obtain the power-law inflationary solution.
Note that an
accelerated expansion is possible 
even if $\alpha>\sqrt{2}$ just as was the case for the U(1) triplet 
electric type inflation\cite{Yamamoto}.

However, the situation in the case of YM field is not exactly the same
as that for the U(1) triplet fields. 
In the above analysis, we have ignored the magnetic component,
which is valid in the  U(1)  triplet  case because the electric and 
magnetic fields are decoupled.
However, the electric and magnetic components are 
always coupled in the YM field. Then 
 we have to check whether the magnetic component is always negligible
 or not when it is initially small.

Since we assume the magnetic component is initially very small,
we can solve the YM equation (\ref{eq_Yang-Mills}) 
dropping the term with $\gYM$ (the magnetic contribution)
as
\begin{eqnarray*}
\dot A&=&A_1 a^{-1}e^{-\lambda\phi}={A_1\over a_0e^{\lambda\phi_0}}
t^{-{\alpha+3\lambda\over 2\alpha}}\,,
\\
A&=&A_0+{2\alpha A_1\over (\alpha-3\lambda) a_0 e^{\lambda\phi_0}}
t^{\alpha-3\lambda\over 2\alpha}
\label{sol_A_2}
\,,
\end{eqnarray*}
where $A_0$ and $A_1$ are integration constants. 
Using this solution,
 we evaluate the ratio of two energy densities as
\begin{eqnarray*}
{\rho_B\over \rho_E}
&\approx
&{\rho_B\over \rho_E}
\Big{|}_0\left({a\over a_0}\right)^4
\,,
\end{eqnarray*}
where ${\rho_B/\rho_E}|_0$ is the initial value. 
We drop $A_0$ since we are interested in the asymptotic behavior 
($t\rightarrow \infty$).
As a result, if the magnetic component is initially sufficiently small,
 we have a power-law inflation just as the case with the Abelian
 multiple fields,
but the contribution of the magnetic component
 gets larger during the evolution of the universe,
and then the inflationary phase eventually ends because of the growth of
the magnetic component.
The e-folding time when the approximation becomes no longer valid 
is evaluated as 
\begin{eqnarray*}
N_{\rm e\mathchar`-folding}&=&\ln(a_{\rm end}/a_0)
\\
&\approx&
-{1\over 4}\ln \left(
{\rho_B\over \rho_E}\Big{|}_0
\right)
\,.
\end{eqnarray*}
For example, if we assume 
 ${\rho_B/\rho_E}|_0=10^{-8}$,
we find $N_{\rm e\mathchar`-folding}\sim 4.6$.
Hence unless the initial value of the magnetic energy density is extremely
small, we do not find a sufficient e-folding number for 
the inflationary universe.

\subsubsection{The case with the dominant magnetic component 
${\rm (}B_{\scriptscriptstyle{\rm YM}}${\rm )}}
We shall call it the regime 
$B_{\scriptscriptstyle{\rm YM}}$, when 
the magnetic component is much larger than the electric one.
In this case the situation is not so simple as the U(1) case because
we cannot ignore the non-linear term in (\ref{eq_YM_density})
even if we assume $\rho_E\ll \rho_B$.
Suppose we have the same solution as the U(1) case.
We then evaluate $\dot A/A$ by the YM equation (\ref{eq_Yang-Mills}),
which is now
\bea
\ddot A+{\alpha+5\lambda\over 2\alpha t}\dot A+2\gYM^2{A^3\over a^2}=0
\,.
\ena
We then find the asymptotic solution as 
\bea
A=A_{\infty}\left[
1+{4\alpha^2 \gYM^2 A_\infty^2\over (\lambda-\alpha)
(\alpha+3\lambda)}{t^2\over a^2}\right]
\,,
\ena
which leads to
\bea
{\dot A\over A}\approx 
{4\alpha \gYM^2 A_\infty\over 
(\alpha+3\lambda)a_0^2}t^{-\lambda/\alpha}
\ena
as $t\rightarrow \infty$. This ratio decays faster
than $t^{-1}$, at which rate $4H \pm \lambda \dot{\phi } $ 
evolve in Eq. (\ref{eq_YM_density}).
So we can ignore the non-linear term with $\dot A/A$,
which gives  exactly the same equations as the $U(1)$ magnetic case.

As a result, 
we have the power-law solution just the same as in 
$B_{\rm U1}$.
This solution is obtained asymptotically, 
and the non-linear term does not destroy it unlike the regime 
$E_{\scriptscriptstyle{\rm YM}}$
where the electric components dominate.

\subsubsection{The case with both components}
If both electric component and magnetic one are of equal magnitude,
we cannot ignore either of them. 
Since it is a non-linearly coupled system, 
it is difficult to figure out
what kind of solutions we expect.
Although we need numerical studies, which we will give later,
here we shall discuss one simple case, in which 
we assume the power-law behavior.

Suppose that the YM potential $A$ is  a power-law function 
as $A\propto t^s$, and the scale factor and the scalar field  
are given by Eqs. (\ref{power-law}) and (\ref{evol_scalar}).
If the energy densities of the electric and magnetic fields are 
similar, i.e., $\rho_E\sim \rho_B$, we find $s=p-1$, and 
then 
\begin{eqnarray*}
\rho_E\sim \rho_B\propto t^{-2}\times t^{-2(1-\lambda/\alpha)}
\,.
\end{eqnarray*}
Inserting this behavior into the YM equation (\ref{eq_Yang-Mills}),
we find
\begin{eqnarray*}
2(p-1)\left(p-1+{\lambda\over \alpha}\right)
+2\gYM^2\left({A_0\over a_0}\right)^2=0
\,,
\end{eqnarray*}
which implies 
\begin{eqnarray*}
2(p-1)\left(p-1+{\lambda\over \alpha}\right)<0
\,.
\end{eqnarray*}
For the power-law inflation ($p>1$), we have
\begin{eqnarray*}
1-{\lambda\over \alpha}>p>1
\,.
\end{eqnarray*}
As a result,
$\rho_E\sim \rho_B$ drops faster than $\propto t^{-2}$, which is 
the scaling of energy density of the scalar field.
Hence the contribution of YM field becomes less 
important as $t\rightarrow \infty$. It appears that it 
is not possible to find a power-law inflation with a
significant residual $\rho _E \sim \rho _B$.
Here, we find the power-law expansion only by a scalar field.
An accelerated expansion is possible 
if $\alpha<\sqrt{2}$, as the conventional power-law inflation.

In the next section, we will 
give more details of interesting solutions 
in the present system 
including the inflationary solutions we have found
and analyze their stability as fixed points in 
a dynamical system.

\section{Stability Analysis}
\label{Stability_Analysis}

\subsection{Dynamical System}
\label{basic_eqs}

In order to analyze the dynamical behavior of our solutions found in \S .
\ref{sol_YM},
we rewrite the basic equations in the form of a  first order autonomous 
system.  The inflationary solutions discussed in the previous 
 sections, along with other interesting ones, appear as fixed points in 
 the dynamical system. This allows us to study their local stability 
 and reveal a complicated dynamical behavior that goes beyond
 the simple power-law time-dependence.
We shall change the time coordinate from $t$ to
the e-folding number $N=\ln (a/a_0)$, and introduce new variables normalized 
by the Hubble expansion rate $H$ as
\bea
{\cal E}&=&e^{{\lambda\over 2}\phi}{E\over H}=
-e^{{\lambda\over 2}\phi}{A'\over a}
\\
{\cal B}&=&e^{{\lambda\over 2}\phi}{B\over H}=
{\cal A}^2
\,,
\\
{\cal A}&=&\gYM^{1/2}e^{{\lambda\over 4}\phi}
{A\over H^{1/2} a}
\,.
\ena
Primes denote differentiations with respect to the e-folding number 
$N$.
We then introduce the density parameter of the 
YM field  as
\bea
\OmYM&=&{\rhoYM\over 3H^2}=\frac{1}{2}\left(\mathcal{E}^2 + 
 \mathcal{B}^2 \right)
\,,
\ena
and  those of the potential and the kinetic energy 
 of the scalar field as
\bea
\Omega_V&=&{V\over 3H^2}
\,,
\\
\Omega_K&=&{\dot\phi^2\over 6H^2}={\varpi^2\over 6}
\,,
\ena
where $\varpi:=\dot \phi/H=\phi'$.
We also use
\bea
\Delta&=&{\rho_B-\rho_E\over \rhoYM}
\,,
\ena
which describes the difference of the fractions
 of the magnetic and electric components. 
 It enables
 a unified treatment of electric- and magnetic-dominant
 regimes and also makes the asymmetry clear when 
 the YM coupling comes into play.
$\Delta=1$ and $-1$ correspond to the regime 
$B_{\scriptscriptstyle{\rm YM}}$ and $E_{\scriptscriptstyle{\rm YM}}$,
respectively.

The Friedmann equation (\ref{Friedmann}) now reads
\bea
\Omega_K +\Omega_V +\OmYM =1
\,.
\label{Friedmann2}
\ena
The equation for the scalar field (\ref{eq_scalar}) is 
\begin{widetext}
\bea
\varpi'={1\over 2}\left(6-\varpi^2\right)
\left(\alpha-\varpi\right)
+\left[2\varpi-3\left(\alpha+\lambda\Delta\right)\right]
\OmYM
\,.
\label{eq_phi:d}
\ena
where we have used the Friedmann equation (\ref{Friedmann2})
to eliminate $\Omega_V$.
The equations for the YM field (\ref{eq_YM_density}) are now
\bea
{\cal A}'&=&{1\over 4}
\left[\varpi(\varpi+\lambda)-4\left(1-\OmYM\right)\right]{\cal A}
-\Gamma{\cal E}
\label{YM_eq1}
\\
{\cal E}'&=&{1\over 2}
\left[\varpi(\varpi-\lambda)-4\left(1-\OmYM\right)\right]{\cal E}
+2\Gamma{\cal A}^3
\,.
\label{YM_eq2}
\ena
where 
\bea
\Gamma=\gYM^{1/2}e^{-{\lambda\over 4}\phi}H^{-1/2}
\,,
\ena
whose evolution equation is 
\bea
\Gamma'&=&{1\over 4}\left[\varpi(\varpi-\lambda)+4\OmYM\right]
\Gamma
\,.
\label{eq_Gamma:d}
\ena
This auxiliary quantity $\Gamma $ is the ``normalized" YM coupling 
in the sense that the subsystem defined by $\Gamma =0$ corresponds
to the dynamical system that describes homogeneous and isotropic
$U(1)$ triplet fields.
 The YM equations (\ref{YM_eq1}) and (\ref{YM_eq2}) are rewritten 
in terms of the denisty parameter $\OmYM$ and the ratio $\Delta$ as
\bea
\OmYM'&=&
\left[-4+\varpi(\varpi+\lambda\Delta)+4\OmYM
\right]\OmYM
\label{eq_YM:d}
\\
\Delta'&=&\lambda\varpi\left(1-\Delta^2\right)-4\epsilon \Gamma
\left(1-\Delta\right)^{1\over 2}\left(1+\Delta\right)^{3\over 4}
\OmYM^{1\over 4}
\,,
\label{eq_delta:d}
\ena
where 
$\epsilon={\rm sign}({\cal A}{\cal E})$.
\end{widetext}
We then find the dynamical system in a closed form.
Since the physical interpretation of the normalized 
vector potential $\mathcal{A}$ is not clear, we take
Eqs.  (\ref{eq_phi:d}),  (\ref{eq_YM:d}), 
(\ref{eq_delta:d}) and (\ref{eq_Gamma:d}) 
with the Hamiltonian constraint 
(= the Friedmann equation) (\ref{Friedmann2}) 
as the basic equations to analyze the stability 
around the fixed points. The drawback is the 
appearance of $\epsilon $ which takes into 
account the ambiguity inherent to taking square 
roots. This causes a problem in the numerical
study in the next section when the system undergoes
oscillations. For this reason, Eqs. (\ref{YM_eq1}) and (\ref{YM_eq2})
instead of Eqs. (\ref{eq_YM:d}) and (\ref{eq_delta:d}) 
are used there.

\subsection{The case with  U(1) triplet  fields}
\label{U(1)}
Before going into analysis of our system, for an introduction and
a comparison, we first summarize 
 the case with the U(1) triplet fields, which 
was discussed in \cite{Yamamoto}, using the present
dynamical variables. To make a clear distinction, we
replace $\Omega _{\rm YM}$ with $\Omega _{\rm U1}$.
Now $\Delta = 1$ and $-1$ correspond to the regimes 
$B_{\rm U1}$ and $E_{\rm U1}$ respectively.

In the case with the U(1) triplet  fields, the dynamical system is 
obtained by setting $\Gamma =0$ in the above;
\begin{eqnarray*}
\varpi'&=&{1\over 2}\left(6-\varpi^2\right)
\left(\alpha-\varpi\right)
\nonumber \\
&&~~+\left[2\varpi-3\left(\alpha+\lambda\Delta\right)\right]
\Omega_{\rm U1}
\\
\Omega_{\rm U1}'&=&
\left[-4+\varpi^2+\lambda\varpi\Delta+4\Omega_{\rm U1}
\right]\Omega_{\rm U1}
\\
\Delta'&=&\lambda\varpi\left(1-\Delta^2\right)
\, .
\end{eqnarray*}

If $\varpi\neq 0$ and $\Omega_{\rm U1}\neq 0$, 
the fixed points are classified into two cases;
$\Delta=-1$ (the case with the electric field) 
and $\Delta=1$ (the case with the magnetic field). 
 In each case,
we find two fixed points as follows:\\
\begin{eqnarray*}
(a) 
&&
\varpi=-3\lambda\Delta\,,~~
\Omega_{\rm U1}={2-3\lambda^2\over 2}\,,
\\
(b) &&
\varpi={4\over \alpha+\lambda\Delta}\,,~~
\Omega_{\rm U1}={\alpha(\alpha+\lambda\Delta)-4\over
 (\alpha+\lambda\Delta)^2}\,.
~~~~
\end{eqnarray*}

Since the density parameters are positive definite, 
$\lambda\leq \sqrt{2/3}$ for the fixed point ($a$) to exist.
In this case, $\Omega_V=0$, which means that
either  the potential is absent from the beginning
or the potential becomes asymptotically
negligible compared with the kinetic term $\varpi^2/2$.

From a perturbative analysis, we can check 
the stability of these fixed points.
For the fixed points ($a$), we find that at least the eigenvalue
for the perturbation of $\Delta$
is always positive ($\omega_\Delta=6\lambda^2$). Hence it is unstable.
Hereafter, we use $\omega $ to denote eigenvalues with subscripts
indicating the variable to which the eigenvalue is associated.

The fixed points ($b$) represent the power-law solutions 
($E_{\rm U1}$ and $B_{\rm U1}$)
found in \S. \ref{power_law_U1}.
The perturbative analysis gives the following three eigenvalues:
\bea
\omega_\Delta=-{8\lambda\Delta\over \alpha+\lambda\Delta}
\ena
and the two roots of the quadratic equation
\begin{eqnarray*}
&&
(\alpha+\lambda\Delta)^2\omega^2+(\alpha+\lambda\Delta)(\alpha+3\lambda\Delta)
\omega
\nonumber 
\\
&&+[\alpha(\alpha+\lambda\Delta)-4][4+3\lambda\Delta(\alpha+\lambda\Delta)]
=0
\,,
\end{eqnarray*}
from the perturbations of $\varpi$ and $\OmYM$.

From the existence conditions given by Eq. (\ref{cond_U1E})
or Eq.(\ref{cond_U1B}), we have $\alpha(\alpha+\lambda\Delta)-4>0$.
Hence, if and only if $\lambda \Delta>0$ is satisfied, 
all the three eigenvalues are negative (or the real parts are negative 
if they are complex).
As a result, the solution with the conditions
\bea
\lambda \Delta>0\,,~~\alpha(\alpha+\lambda\Delta)-4> 0
\ena
is stable against linear perturbations.
More concretely, the stability conditions 
for the cases with the electric field 
($E_{\rm U1}$) and magnetic field ($B_{\rm U1}$)
are given by, 
\bea
&&
\lambda<0\,,~~\alpha(\alpha-\lambda)>4
\,,
\\
&&
\lambda >0\,,~~\alpha(\alpha+\lambda)>4
\,,
\ena
respectively. If $\lambda>0$, the   magnetic  power-law solution
 (Eqs. (\ref{power-law}), 
 (\ref{evol_scalar}) and (\ref{sol_EM}) with (\ref{U1B_p}) and (\ref{U1B_B}) )
 is always an attractor in the parameter range of $
\alpha(\alpha+\lambda)>4$, while if  $\lambda<0$
the  electric power-law solution (Eqs. (\ref{power-law}), 
 (\ref{evol_scalar}) and (\ref{sol_EM}) with (\ref{U1E_p}) and (\ref{U1E_E}) )
 is always an attractor in the parameter range of $
\alpha(\alpha-\lambda)>4$.

For the rest of the parameter space 
($\alpha-4/\alpha<\lambda<-\alpha+4/\alpha$), 
the attractor is a fixed point with $\Omega_{\rm U1}=0$,
where the scalar field dominates the universe.
The fixed point, which we denote $S_{\rm U1}$, is given by
\bea
\varpi=\alpha\,,~~\Omega_{\rm U1}=0\,,~~\Delta=\pm 1
\,.
\ena
The perturbative analysis gives
the three eigenvalues as 
\begin{eqnarray*}
\omega_\varpi&=&-{1\over 2}(6-\alpha^2)\,,
\\
\omega_{\Omega_{\rm U1}}&=&\alpha(\alpha+\lambda\Delta)-4
\,,
\\
\omega_\Delta&=&-2\alpha\lambda\Delta
\,.
\end{eqnarray*}
Hence the power-law solution driven only by the scalar field 
is stable if $\lambda\Delta>0$ and $\alpha(\alpha+\lambda\Delta)-4<0$.
Between the two solutions with alternative sings, the stable one is 
\bea
\varpi=\alpha\,,~~\Omega_{\rm U1}=0\,,~~\Delta=1
\,.
\ena
for $\lambda>0$, while 
\bea
\varpi=\alpha\,,~~\Omega_{\rm U1}=0\,,~~\Delta=-1
\,.
\ena
for $\lambda<0$.

\begin{widetext}
We summarize the result for the U(1) triplet case
in Fig. \ref{fig:parameter_U1} and 
 in Table \ref{table1}:
\begin{figure}[h]
\begin{center}
\includegraphics[scale=.6]{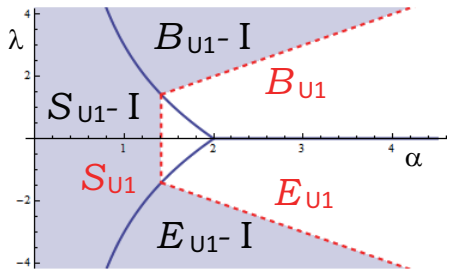}
\caption{
The parameter range for power-law solutions
in the case with the  U(1)  triplet fields. 
 The inflationary attractors are indicated by black letters
while non-inflationary attractors are red.
The attractor solution with the electric field is
given for $\lambda<0$ and $\alpha(\alpha-\lambda)>4$
($E_{\rm U1}$),
while one with the magnetic field is for  
$\lambda>0$ and $\alpha(\alpha+\lambda)>4$ ($B_{\rm U1}$) .
In the range of $\alpha-4/\alpha<\lambda<-\alpha+4/\alpha$
($S_{\rm U1}$),
we find the attractor dominated by the scalar field.
All inflationary solutions 
 are stable.
The inflation with the  U(1)  triplet field 
is found in the range either 
of $\lambda<-\alpha$ and $\alpha(\alpha-\lambda)>4$
($E_{\rm U1}$-I:with the electric field)
or of  $\lambda>\alpha$ and $\alpha(\alpha+\lambda)>4$
($B_{\rm U1}$-I:with the magnetic field).
The conventional power law inflation 
with an exponential potential is possible 
only for $\alpha<\sqrt{2}$ and
$\alpha-4/\alpha<\lambda<-\alpha+4/\alpha$.
($S_{\rm U1}$-I)}
\label{fig:parameter_U1}
\end{center}
\end{figure}

~~\\[-3em]
\begin{center}
\begin{table}[h]
\begin{tabular}{|c||c|c|c|}
\hline
\hline
fixed point&$S_{\rm U1}$&$E_{\rm U1}$&$B_{\rm U1}$
\\
\hline
\hline
existence & $\alpha <\sqrt{6}$ 
&$\alpha-{4/\alpha}>\lambda $ 
&$-\alpha+{4/\alpha}<\lambda $
\\[.2em]
\hline
\\
~&~&~&~
\\[-2.2em]
$p$ &${2/\alpha^2}$&
${1\over 2}\left(1-{\lambda/\alpha}\right)$
&${1\over 2}\left(1+{\lambda/\alpha}\right)$
\\[.2em]
\hline
\\
~&~&~&~
\\[-2.2em]
$\OmYM$&$0$ &${\alpha(\alpha-\lambda)-4\over (\alpha-\lambda)^2}$ &
${\alpha(\alpha+\lambda)-4\over (\alpha+\lambda)^2}$
\\[.2em]
\hline
\\
~&~&~&~
\\[-2.2em]
$\Omega_{V}$&$1-{\alpha^2/6}$ 
&${4-3\lambda(\alpha-\lambda)\over 3(\alpha-\lambda)^2}$ 
&${4+3\lambda(\alpha+\lambda)\over 3(\alpha+\lambda)^2}$ 
\\[.2em]
\hline
\\
~&~&~&~
\\[-2.2em]
$\Omega_{\rm K}$&${\alpha^2/6}$ &
${8\over 3(\alpha-\lambda)^2}$ 
&${8\over 3(\alpha+\lambda)^2}$
\\[.2em]
\hline
\\
~&~&~&~
\\[-2.2em]
stability
&$\alpha-{4/\alpha}<\lambda<-\alpha+{4/\alpha}$  
& $\lambda<0$
& $\lambda>0$
\\
\hline
\hline
inflation &   $\alpha<\sqrt{2}$~~~~~~~($S_{\rm U1}$-I)
&  $\lambda<-\alpha$ ~~($E_{\rm U1}$-I)
&   $\lambda>\alpha$ ~~($B_{\rm U1}$-I)
\\
\hline
\end{tabular} 
\caption{The fixed points and their properties for the case
 with U(1) triplet fields.  The second row gives the existence conditions.
The bottom two rows are
 understood to hold when those existence conditions are satisfied.}
\label{table1}
\end{table}
\end{center}
\end{widetext}

\subsection{Important Fixed Points in the dynamical system with YM field}
\label{YM_fixed points}
Now we move on to include the non-linear YM interaction.
The non-trivial fixed points are classified into two cases:
$\Gamma=0$ and $\Gamma\neq 0$.
In the former case, we find the same fixed points as the  U(1)  triplet case,
although their stability is different as we will show later.
The latter case gives new fixed points, which do not exist in the  
U(1) triplet system.

Note that  a fixed point may not be found by 
an exact solution, but can be reached as a certain limit.
For example,  the fixed points with $\Gamma=0$ would
imply either $g_{\rm YM}=0$ or $H e^{\lambda \phi /2} = \infty $, 
neither of which is of interest in our analysis here. However, 
starting from $g_{\rm YM} \neq 0 $ and finite $H$ and $\phi $, 
the system may approach $\Gamma \rightarrow 0$ asymptotically 
as $t\rightarrow \infty$. From the mathematical point of view,
those fixed points are well-defined and a part of the dynamical system 
and we include them in the following analysis.

\subsubsection{$\Gamma=0$}

In this case,  which should reproduce the 
fixed points of the previous subsection, we 
can classify the solutions into two cases:
$\OmYM=0$ and $\OmYM\neq 0$.
\\

\underline{{\bf (a)} $\OmYM=0$}\\[.5em]
In the case with $\OmYM=0$, the scalar field energy is dominant.
From (\ref{eq_phi:d}), we find either $\varpi^2=6$ or $\varpi=\alpha$ with 
$\Delta^2=1$.
The former fixed point corresponds to the case that the kinetic energy 
of the scalar field is dominant, 
which is unstable against perturbations. The latter fixed points denote
the power-law expanding universe with an exponential potential
 (the counterpart of $S_{\rm U1}$) and will be called $S_{\rm YM}$.
The ratio of  the potential energy $V$ to the kinetic energy  is
$(6-\alpha^2)/\alpha^2$. As is well known, these fixed points are attractors 
if $\alpha\leq \sqrt{6}$ for the case only with a scalar field.

In the present case, because of the YM field, 
the stability condition changes as follows.
The linearized equations for these fixed points
 give four eigenvalues:
\bea
\omega_{\varpi}&=&{1\over 2}\left(\alpha^2-6\right)
\,,
\label{om_varpi}
\\
\omega_{\OmYM}&=&\alpha\lambda\Delta+\alpha^2-4
\,,
\label{om_OmYM}
\\
\omega_{\Delta}&=&-2\alpha\lambda\Delta
\,,
\label{om_Delta}
\\
\omega_{\Gamma}&=&{1\over 2}\alpha(\alpha-\lambda)
\, .
\label{om_Gamma}
\ena
Three eigenvalues (\ref{om_varpi}), (\ref{om_OmYM}), and (\ref{om_Delta}) 
are all negative if $\lambda>0$ and $\lambda<-\alpha+4/\alpha$ for $\Delta=1$,
or if $\lambda<0$ and $\lambda >\alpha-4/\alpha$ for $\Delta=-1$.
The forth eigenvalue (\ref{om_Gamma}) becomes negative if $\lambda>\alpha$.
As a result, the fixed point with  $\Delta=1$ is stable in the 
parameter range of $\alpha<\lambda<-\alpha+4/\alpha$.

On the other hand,  taking $\lambda < \alpha $ gives instability against 
the perturbations of $\Gamma$. However, as long as $\Omega _{\rm YM}$
stays small, the growing $\Gamma $ does not disturb the evolution of the universe
as well as the dynamics of the scalar field since $\Gamma $ does not appear
explicitly in Eqs. (\ref{Friedmann2}) and (\ref{eq_phi:d}). This is indeed the 
case when $\alpha-4/\alpha<\lambda<\alpha $. As we shall confirm later 
in the numerical analysis, the dynamics of the universe is dominated by
the scalar field and accurately described by the fixed point discussed here,
despite the apparent instability in the eigenvalue $\omega _{\Gamma }$.
The exponentially increasing $\Gamma $ only triggers a rapid oscillation
for the perturbed YM field whose amplitude remains small.

In summary, we conclude that these fixed points are stable in the range
$\alpha -4/\alpha < \lambda < -\alpha +4/ \alpha $ as was found for 
$S_{\rm U1}$. Nevertheless, there is a distinction between $S_{\rm U1}$
and $S_{\rm YM}$ for $\lambda< \alpha $
with the dynamics of the YM field being different.
\\

\underline{{\bf (b)} $\OmYM\neq 0$}\\[.5em]
For the case with $\OmYM\neq 0$, the YM field plays an important 
role in the dynamics of the universe.
We find $\Delta=\pm 1$ unless $\varpi=0$, for which we do not 
have any interesting dynamics.

$\Delta=-1$ and $1$ correspond to 
the case of the electric component dominance 
($E_{\scriptscriptstyle{\rm YM}}$) and that of
the magnetic component dominance ($B_{\scriptscriptstyle{\rm YM}}$), 
respectively.
As we have already mentioned, the YM field always 
consists of both components.
Hence these fixed points are reached only asymptotically, 
if they are stable.
Just the same as the U(1) triplet fields, we find two fixed points 
for each case. However, one of them with $\varpi=- 3\lambda\Delta$ is unstable.
Hence we discuss the other cases:
\bea
(1)
&&\Delta=1~~ (B_{\scriptscriptstyle{\rm YM}})
\nonumber \\
&&
~~\varpi={4\over \alpha+\lambda}\,,~~
\nonumber \\
&&
~~
\OmYM={\alpha(\alpha+\lambda)-4\over (\alpha+\lambda)^2}\,.
\\
(2) 
&&\Delta=-1~~ (E_{\scriptscriptstyle{\rm YM}})
\nonumber \\
&&~~\varpi={4\over \alpha-\lambda}\,,~~
\nonumber \\
&&
~~\OmYM={\alpha(\alpha-\lambda)-4\over (\alpha-\lambda)^2}\,.
~~~~~~
\ena
These fixed points correspond to the solutions with 
the magnetic component and the electric one
 found in \S. \ref{MU1} and \ref{EU1}, respectively.

Without the non-linear interaction, these points were symmetric:
they were related by the electromagnetic duality and had the 
same stability properties. 
The YM coupling skews the symmetry.

For the case (1), the eigenvalues are given by
\bea
\omega_\Delta&=&-{8\lambda\over \alpha+\lambda}
\\
\omega_\Gamma&=&{2(\alpha-\lambda)\over \alpha+\lambda}
\ena
and the two roots of 
the algebraic equation
\bea
&&
(\alpha+\lambda)^2\omega^2+(\alpha+\lambda)(\alpha+3\lambda)\omega
\nonumber \\
&&~~+[\alpha(\alpha+\lambda)-4][4+3\lambda(\alpha+\lambda)]=0
\,.~~~~~
\ena
We find all eigenvalues are negative if and only if 
$\lambda>\alpha$ and $\alpha(\alpha+\lambda)-4>0$ are satisfied.
Since this condition corresponds to the power-law inflationary solution
with YM field, 
we can conclude that the power-law inflationary solution 
with magnetic component dominance ($B_{\rm YM}$-{\rm I}) is an attractor.
The difference from the U(1) multiplet case is that the solutions in the 
parameter range of
$0<\lambda<\alpha$, which are not inflationary, are no longer an attractor.
We will discuss later which asymptotic state we find in this region. 

On the other hand, in the case with $\Delta=-1$,
the eigenvalues are given by
\bea
\omega_\Delta&=&{8\lambda\over \alpha-\lambda}
\\
\omega_\Gamma&=&2
\ena
and the two roots of 
the algebraic equation
\bea
&&
(\alpha-\lambda)^2\omega^2+(\alpha-\lambda)(\alpha-3\lambda)\omega
\nonumber \\
&&~~+[\alpha(\alpha-\lambda)-4][4+3\lambda(\alpha-\lambda)]=0
\,.~~~~~
\ena
We find that three eigenvalues are negative for the power-law 
inflationary solution as long as 
$\lambda<-\alpha$ and $\alpha(\alpha-\lambda)-4>0$,
but one eigenvalue $\omega_\Gamma$, which corresponds to
the perturbations of $\Gamma$ (non-linear interaction term of the YM field),
 is always positive and
 does not depend on any parameters. 
This is the same behavior which we have seen in \S.\ref{EYM}.
As a result, this solution is
unstable and the typical instability time scale is O(1) e-folding
time since the present time coordinate is $N=\ln (a/a_0)$ and
therefore $\Gamma \propto \exp \left( \omega _{\Gamma }N \right)$.
If the magnetic component is initially sufficiently small,
we may find this power-law inflation solution by the electric components
in the beginning, 
but the orbit leaves it just after $O(1)$ e-folding time. 
We conclude that the power-law inflationary solution 
with the electric component dominance $E_{\rm YM}$-I
is unstable, contrary to the $E_{\rm U1}$.

\subsubsection{$\Gamma\neq 0$}

Here we assume that $\OmYM\neq 0$ because $\Gamma$ 
becomes important when the YM field gives non-trivial contribution 
to the cosmic expansion.
Note that $\Delta^2=1$ ($E_{\scriptscriptstyle{\rm YM}}$ 
or $B_{\scriptscriptstyle{\rm YM}}$) is no longer a fixed point.
We find the following two non-trivial fixed points, if $\lambda^2\geq 6$:
\begin{eqnarray*}
\varpi=\varpi^{(\pm)}:={3\over 2}\left(\lambda\pm \sqrt{\lambda^2-16/3}\right)
\end{eqnarray*}
The values at both fixed points can be described neatly by
 the deceleration parameter 
\begin{eqnarray}
q^{(\pm)}:=1+{\left(\varpi^{(\pm)}\right)^2\over 6}
={\lambda\varpi^{(\pm)}\over 2}-1
\label{sol_q}
\,,
\end{eqnarray}
as 
\begin{eqnarray*}
&&
\OmYM^{(\pm)}=2-q^{(\pm)}\,,~~
\Delta^{(\pm)}={1-q^{(\pm)}\over 1+q^{(\pm)}}\,,~~
\nonumber \\
&&
\Gamma^{(\pm)}=\left[{(q^{(\pm)})^2(1+q^{(\pm)})\over 2(2-q^{(\pm)})}
\right]^{1/4}
\,,
\nonumber \\
&&
p^{(\pm)}={1\over q^{(\pm)}+1}
\,.
\end{eqnarray*}
Note that $\Omega_V=0$ at these fixed points
 and they are unique to the YM case.
From (\ref{sol_q}) and 
the positivity of the density parameter $\OmYM\geq 0$,
we find
\begin{eqnarray}
1\leq q^{(\pm)}\leq 2
\,.
\label{range_q}
\end{eqnarray}
Hence the power exponent of the scale factor $p^{(\pm)}$ is
\begin{eqnarray*}
1/3\leq p^{(\pm)}\leq 1/2
\,,
\end{eqnarray*}
which is between a radiation dominant state and a stiff-matter dominant 
one.

The perturbative analysis is common to both of the 
fixed points $\lambda\geq \sqrt{6}$ ($NA_+$) and
$\lambda\leq -\sqrt{6}$ ($NA_-$).
We find the following eigenvalues;
\begin{eqnarray*}
\omega_{\OmYM\mathchar`-\varpi}&=&
2\left(1-{\alpha\over \lambda}\right)(q+1)
\,,
\\
\omega_{\Gamma\mathchar`-\Delta}&=&q-2
\,,
\end{eqnarray*}
which are associated with the eigen vectors 
$\delta\OmYM+(\varpi/3)\delta\varpi$ and 
$\delta\Gamma/\Gamma-
[(q-2)(q+1)/8q] \delta \Delta$, respectively, 
and the two roots of the quadratic equation
\begin{eqnarray*}
\omega^2+(2-q)\omega+{4q(3-q)\over q-1}=0
\,.
\end{eqnarray*}
Since $q$ is in the range of (\ref{range_q}), if $0<\lambda<\alpha$, we find 
all eigenvalues are negative,   which means $NA_+$ is stable.
On the other hand, $\omega _{\Omega _{\rm YM} \mathchar`- \varpi}$
turns positive when $\lambda <0$ and we find $NA_-$ is unstable.
As a result, we find a stable fixed point in the parameter range of 
 $\sqrt{6}<\lambda<\alpha$ ($NA_+$),  which partly takes care of 
 the lost stability of $B_{\rm YM}$ in the region $\lambda <\alpha $.

We summarize our result for the SU(2) YM field
in Fig. \ref{fig:parameter_YM} and 
 in Table \ref{table2}:
\begin{widetext}
~
\begin{figure}[h]
\begin{center}
\includegraphics[scale=.5]{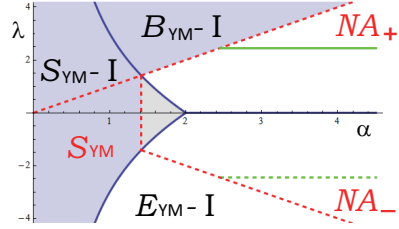}
\caption{
The parameter range for power-law solutions
in the case with the YM field. 
The inflationary attractor solution with  the magnetic component  
is found for  
$\alpha<\lambda<-\alpha+4/\alpha$
($\BYM$-I).
On the other hand, the inflationary solution with
the electric component, which is
found in the range of  $\alpha-4/\alpha<\lambda<-\alpha$
($\EYM$-I),
is unstable.
In the range of $\alpha-4/\alpha<\lambda<-\alpha+4/\alpha$
($\SYM$),
we find the attractor solutions dominated by a scalar field.
For inflation, we need an additional condition
$\alpha<\sqrt{2}$ ($\SYM$-I).
We also find new fixed points $NA_\pm$, which exist 
only for non-Abelian gauge fields.}
\label{fig:parameter_YM}
\end{center}
\end{figure}

\begin{center}
\begin{table}[h]
\begin{tabular}{|c||c|c|c|c|c|}
\hline
\hline
&$\SYM$&$\EYM$&$\BYM$&$NA_+$&$NA_-$
\\
\hline
\\
~&~&~&~&~&~
\\[-2.2em]
existence &{\scriptsize $ \alpha < \sqrt{6} $}
&{\scriptsize $ \alpha -4/\alpha > \lambda  $}
&{\scriptsize $\lambda> - \alpha+{4/\alpha}$}
&{\scriptsize $\sqrt{6}<\lambda $}
&{\scriptsize $ \lambda <-\sqrt{6} $}
\\[.2em]
\hline
\\
~&~&~&~&~&~
\\[-2.2em]
$p$ &{\scriptsize ${2/\alpha^2}$}&
{\scriptsize ${1\over 2}\left(1-{\lambda/\alpha}\right)$}
&{\scriptsize ${1\over 2}\left(1+{\lambda/\alpha}\right)$}
&{\scriptsize ${1\over 4\lambda}(\lambda+\sqrt{\lambda^2-16/3})$}
&{\scriptsize ${1\over 4\lambda}(\lambda-\sqrt{\lambda^2-16/3})$}
\\[.2em]
\hline
\\
~&~&~&~&~&~
\\[-2.2em]
$\OmYM$&$0$ &{\scriptsize
 ${\alpha(\alpha-\lambda)-4\over (\alpha-\lambda)^2}$} &
{\scriptsize ${\alpha(\alpha+\lambda)-4\over (\alpha+\lambda)^2}$}
&{\scriptsize ${3\over 4}(4-\lambda^2+\lambda\sqrt{\lambda^2-16/3})$}
&{\scriptsize ${3\over 4}(4-\lambda^2-\lambda\sqrt{\lambda^2-16/3})$}
\\[.2em]
\hline
\\
~&~&~&~&~&~
\\[-2.2em]
$\Omega_{V}$&{\scriptsize $1-{\alpha^2/6}$ }
&{\scriptsize ${4-3\lambda(\alpha-\lambda)\over 3(\alpha-\lambda)^2}$ }
&{\scriptsize ${4+3\lambda(\alpha+\lambda)\over 3(\alpha+\lambda)^2}$ }
&0&0
\\[.2em]
\hline
\\
~&~&~&~&~&~
\\[-2.2em]
$\Omega_{\rm K}$&{\scriptsize ${\alpha^2/6}$} &
${8\over 3(\alpha-\lambda)^2}$ 
&${8\over 3(\alpha+\lambda)^2}$
&{\scriptsize ${3\over 4}(\lambda^2-8/3-\lambda\sqrt{\lambda^2-16/3})$}
&{\scriptsize ${3\over 4}(\lambda^2-8/3+\lambda\sqrt{\lambda^2-16/3})$}
\\[.2em]
\hline
stability
& {\scriptsize $\alpha -4/\alpha < \lambda < -\alpha +4/\alpha $}
& always unstable
& {\scriptsize $\alpha < \lambda $}
&{\scriptsize $ \lambda < \alpha $} 
& always unstable
\\
\hline
\hline
\end{tabular} 
\caption{The fixed points and their properties for the case
 with YM fields. Stability of $S_{\rm YM}$ takes into account the
 fact that unstable $\Gamma $ does not destroy the dominance 
 of the scalar field. $NA_{\pm }$ cannot be inflationary. Inflationary
 conditions for the other points are the same as the $U(1)$ triplet case.}
\label{table2}
\end{table}
\end{center}

\section{Numerical Study}
\label{Numerical_Study}

From the above stability analysis, we find 
there are stable attractors if $\lambda\geq \alpha$ 
($B_{\scriptscriptstyle{\rm YM}}$-I)
or $\alpha \geq \lambda \geq \sqrt{6}$ ($NA_+$).
We also find that a scalar field dominated universe
($S_{\scriptscriptstyle{\rm YM}}$),
which is the same as the stable attractor in the model with 
a scalar field with an exponential potential ($V=V_0
\exp(-\alpha\phi)$), is stable in the parameter range of 
$\alpha-4/\alpha \leq \lambda\leq -\alpha+4/\alpha$,
 even though the YM field does not necessarily settle
down to its attractor state.
As we will show here, it will oscillate in this scalar
dominated background.

We may also wonder what is the future asymptotic
behavior for the other range of the coupling parameters
 $\alpha$ and $\lambda$, i.e., 
$\lambda>\alpha$ and $\lambda<\sqrt{6}$.
Numerical calculations give us some insight 
into this question too.

Numerical study also tells us 
strengths of the stable attractors.
Since our stability analysis is based on the
 linear perturbations, we need numerical analysis 
to know how the attractor state is achieved from 
generic initial data.
\end{widetext}

~~\\
\vskip 4em
\subsection{Numerical Analysis}
\subsubsection{Stable attractors}
We begin with a small value of $\alpha $ for which the conventional power-law
 inflation is known 
to occur in the absence of gauge-kinetic coupling, namely $\alpha <\sqrt{2}$.

\begin{figure}[h]
\begin{center}
\includegraphics[scale=0.7]{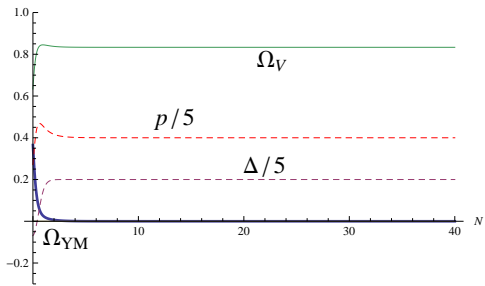}
\caption{Inflation for $\alpha =1 , \lambda =2$. This case obeys 
the usual cosmic-no-hair. }
\label{fig:powerB}
\end{center}
\end{figure}

 We choose 
the representative value to be $\alpha =1$. 
We first performed the calculation for $\lambda=2$
(see Fig. \ref{fig:powerB}),
which shows the conventional power-law inflation 
with an exponential potential ($\SYM$-I ). The YM field energy 
drops quickly.
We find that the asymptotic power exponent of the scale factor is 2,
which is consistent with the value of
 the conventional power-law inflation ($p=2/\alpha^2$).

\begin{figure}[h]
\begin{center}
\includegraphics[scale=0.7]{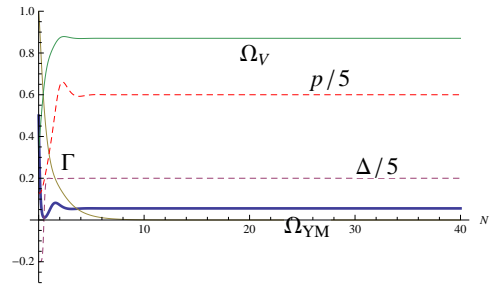}
\caption{Occurrence of inflation assisted by the magnetic field for $\alpha = 1,
 \lambda =5$. Convergence to the inflating attractor $\BYM$-I 
 is clearly seen. }
\label{fig:mag1}
\end{center}
\end{figure}

When $\lambda > 3$, our analysis 
suggests the power-law inflation assisted by the magnetic component of the YM 
field ($\BYM$-I) is a stable attractor of the system. 
Fig.\ref{fig:mag1} confirms this fact as the density parameter for magnetic 
component stays constant ($\OmYM$=constant and $\Delta=1$)
while the scalar potential dominates the energy budget, which implies the 
universe undergoes accelerated expansion.  An important difference
between Figs.\ref{fig:powerB} and \ref{fig:mag1} is that the 
acceleration is actually stronger when $\Omega _{\rm YM}$ 
does not vanish. We indeed find the asymptotic value of
the power exponent $p$ is 3 instead of 2.
We deliberately chose the initial condition such that the scalar kinetic 
energy and the electric component 
are dominant over the others and the effect of YM coupling is  
significant. 
As shown in Fig.\ref{fig:mag1}, $\Delta$ approaches unity and 
$\Gamma$ decays quickly whereby the system essentially 
reduces to the $U(1)$ triplet model.
We find $\BYM$-I asymptotically.

\begin{figure}[h]
\begin{center}
\includegraphics[scale=0.7]{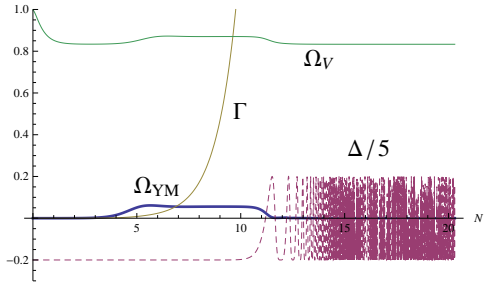}
\caption{Inflation $\SYM$-I with electromagnetic oscillation via the electric 
YM inflation ($\EYM$-I) for $\alpha = 1, \lambda =-5$. The initial value of 
the ratio of the energy density of the magnetic component to 
that of the electric one is $10^{-8}$.
Between $5<N<10$, 
$\Omega_V$ is greater than its final value, which means the acceleration is
 stronger during that period thanks to the help by the electric component. 
As $\Gamma $ gets to order unity, this regime is ruined and 
$\OmYM$ decays while the fields oscillate.}
\label{fig:ele1}
\end{center}
\end{figure}

Next, we take a negative $\lambda $ and confirm the electric-magnetic
 asymmetry for 
non-Abelian gauge fields. Fig.\ref{fig:ele1} exhibits two different regimes. 
In the beginning, 
the electric energy density grows according to the linear instability caused
 by the strong
gauge-kinetic coupling and the system is attracted towards the  power-law
inflation assisted by electric component of the YM field ($\EYM$-I). 

During that period, however, $\Gamma $ continues to increase and eventually
 destroys the 
inflationary regime at $N \sim 10$. 
The transient inflation $\EYM$-I continues for $5\sim 6$ e-folding 
number, which is consistent with our evaluation given in \S. \ref{EYM}.

\begin{figure}[h]
\begin{center}
\includegraphics[scale=0.7]{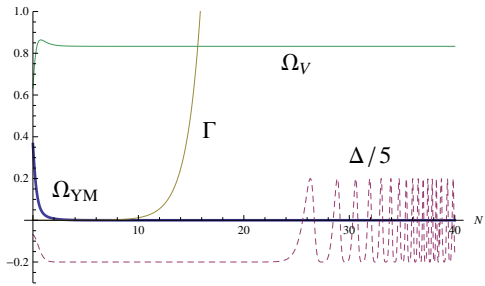}
\caption{Inflation for $\alpha =1 , \lambda =-2$. While the dynamics of the universe is entirely dominated by the scalar field, gauge fields oscillate at late time.}
\label{fig:powerE}
\end{center}
\end{figure}

After that, the universe is dominated
 by the scalar field
while YM field is oscillating. In this case, since
 $\alpha $ is small enough to cause accelerated
expansion by itself, this oscillation phase is also inflating. For comparison,
 we also show the plots with a smaller value of $|\lambda |$ 
(Fig.\ref{fig:powerE}). 
The behavior is similar to Fig. \ref{fig:ele1}, but
there is no transient regime of $\EYM$-I.
When $\lambda $ is negative, from the instability of $\EYM$-I, 
there is a peculiar behavior of rapid oscillation at late time 
between electric  and magnetic components,
 which is not seen for positive $\lambda $.

\begin{figure}[h]
\begin{center}
\includegraphics[scale=0.7]{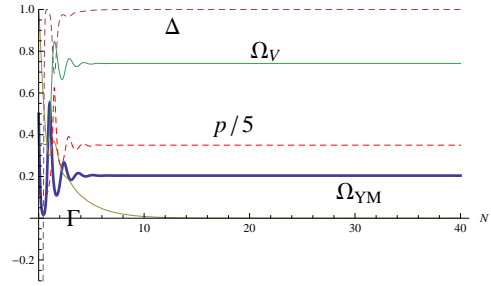}
\caption{Inflation assisted by the YM field ($\BYM$-I) for $\alpha =2 , \lambda =5$.
Note that the asymptotic value of $\Omega _V$ is sufficiently large to cause 
accelerated expansion. Although the slope of the scalar potential is not flat 
enough to maintain the potential domination by itself, the magnetic 
component of the YM field 
also takes up the scalar field energy and helps realizing the inflation.}
\label{fig:mag2}
\end{center}
\end{figure}

Let us turn our attention to the supportive role of gauge fields in realizing 
inflation. We take $\alpha =2$ for which inflation is impossible by the scalar
 field itself. With $\lambda =5$, we obtain Fig. \ref{fig:mag2} where 
$\Delta = 1$ in the future
 asymptotic state. The value of $\Omega _V$ close to unity shows the expansion 
is accelerated, which can also be seen by the power exponent $p>1$. 
 As was investigated in the previous sections, this is due to the interaction
 between the scalar and YM fields that transfers scalar field energy to 
magnetic component of the YM field and slows down its rolling down
 the potential.

Note that the velocity of the scalar field is given by 
$\dot \phi=2\ln t /\alpha$, which is the same as the conventional 
power-law inflation. The difference is the values of total energy densities.
The effective potential in the present model is given by 
the YM energy as well as the scalar potential $V$ (Eq. (\ref{potential_eff})), 
which gives a larger Hubble expansion rate. As a result,
the velocity with respect to the e-folding number $N$ 
becomes slower as $\phi'=\dot\phi/H$.

\begin{figure}[h]
\begin{center}
\includegraphics[scale=0.7]{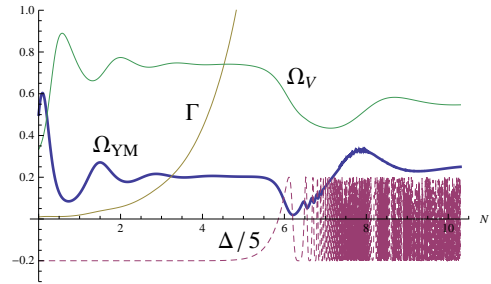}
\caption{Inflation for $\alpha =2 , \lambda =-5$. Intermediately 
($ 2 \lesssim  N \lesssim 6$), the universe briefly inflates ($\EYM$-I). Then $\Gamma $
 eventually dictates the dynamics of the YM field.}
\label{fig:noinfE}
\end{center}
\end{figure}
For the $U(1)$ gauge fields, the same type of inflation with non-flat 
potential $E_{\rm U1}$-I could have been seen for negative $\lambda $ 
because of the electro-magnetic duality. In the present non-Abelian case, 
however, a negative $\lambda $ drives not only electric component but also the
 normalized gauge coupling 
$\Gamma $, by which the inflationary regime is made transient and the 
final state contains mixture 
of electric, magnetic and scalar fields (Fig.\ref{fig:noinfE}). During 
the transient phase of inflation 
supported by the electric component of the YM field ($\EYM$-I), 
one can see the values of $\Omega _V $ and $\OmYM$ 
being the same as the corresponding magnetic inflation.

\begin{figure}[h]
\begin{center}
\includegraphics[scale=0.7]{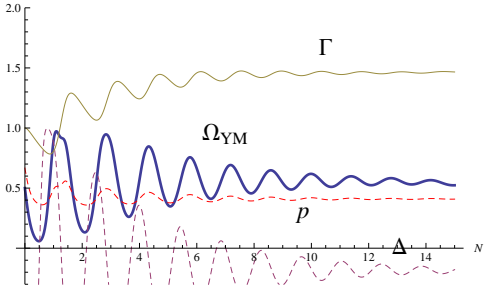}
\caption{Convergence to the non-Abelian attractor 
$NA_+$ for $\alpha =4 , \lambda =3$. 
It is distinct from the other plots in that $\Gamma $ settles down to 
a constant value. }
\label{fig:osc2}
\end{center}
\end{figure}

Finally, we confirm the stability of the new non-Abelian fixed point
$NA_+$ for $\lambda > \sqrt{6}$ (Fig.\ref{fig:osc2}). 
The convergence is relatively slow and all the dynamical components 
undergo oscillations. In contrast to 
the other cases where $\Gamma $ either diverges or dies away, 
this parameter region sees convergence
to an attractor value, which necessarily means negligible $\Omega _V$. 
Although it is not of interest in
the context of inflation, it illustrates a distinct effect of the gauge 
coupling by forcing the potential term to vanish that would never happen
 in scalar-U(1) systems.

\subsubsection{Oscillation of the YM field}

The focus of this subsection is to understand the future asymptotic behavior of 
the system in the parameter region where the elementary fixed point analysis 
suggests there is no stable attractor solution. It turns out the nature of 
the dynamics in this regime is oscillation driven by the gauge coupling.

\begin{figure}[h]
\begin{center}
\includegraphics[scale=0.7]{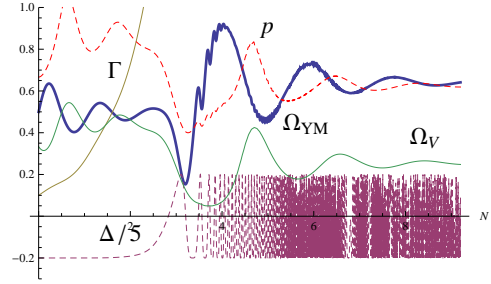}
\caption{Oscillation for $\alpha =4 , \lambda =-3$. As is expected, while 
$\Gamma $ is smaller than unity, the system approaches the  
fixed point with the electric component 
$\EYM$. After the effect of non-linear gauge coupling kicks in, 
the dynamics is irregular at the beginning. It appears the oscillation of 
$\Omega _V$ and $\OmYM$ eventually die away, finding 
some asymptotic solution with the YM field oscillations. 
The power exponent $p$ of the scale factor is slightly larger 
than $1/2$.}
\label{fig:osc1}
\end{center}
\end{figure}

Fig.\ref{fig:osc1} shows the occurrence of scalar-YM oscillation as the 
future asymptotic state of the dynamical system for $\alpha =4, \lambda =-3$. 
As is
 expected, the potential energy does not play a prominent role here.
 $\Omega _V$ and $\OmYM$ 
 appear to converge to finite values although the numerical calculation 
has not been able to 
 confirm it due to the computational difficulty caused by the rapid 
oscillation of $\Delta $ and
 the ever-growing $\Gamma $. 

\begin{figure}[h]
\begin{center}
\includegraphics[scale=0.7]{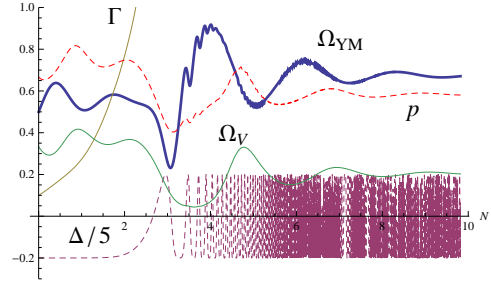}
\caption{Oscillation for $\alpha =4, \lambda =-2$. Qualitatively the same 
dynamics as $\lambda =-3$. The only essential difference from 
Fig.\ref{fig:osc1} is the asymptotic value of $\Omega _V$ and $\OmYM$.}
\label{fig:osc3}
\end{center}
\end{figure}
The behavior is mostly the same for negative $\lambda $ regardless of 
$\lambda < -\sqrt{6}$ or not (Fig.\ref{fig:osc3}). 
The power exponent $p$ of the scale factor is always slightly larger 
than $1/2$.

\begin{figure}[h]
\begin{center}
\includegraphics[scale=0.7]{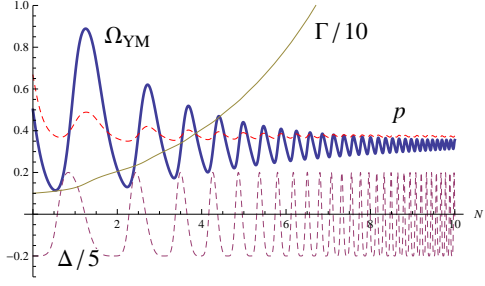}
\caption{Oscillation for $\alpha =4 , \lambda =2$. One can see the 
exponentially decaying amplitude of oscillation for $\OmYM$. 
In contrast to $\lambda <0$, the scalar potential contribution
becomes completely negligible.}
\label{fig:osc4}
\end{center}
\end{figure}
In contrast, $\lambda =\sqrt{6}$ is a threshold 
value for positive $\lambda $ since 
the YM fixed point becomes the attractor above it (see Fig.\ref{fig:osc2}). 
Below the critical value, the asymptotic dynamics is rather analogous to 
the cases with negative $\lambda $, 
but with a significantly smaller contribution of $\Omega _V$. 
Convergence to an 
asymptotic value for $\OmYM$ can be seen more clearly here 
(Fig.\ref{fig:osc4}).
The power exponent $p$ of the scale factor is slightly smaller 
than $1/2$ for $\lambda>0$.

\subsection{Asymptotic spacetime with the oscillation of the YM field} \label{sec:iteration}

From our numerical study, we find that the universe still
approaches some attractor spacetime but the YM field is oscillating
for some parameter range, where we do not find stable attractors.
In order to identify such an attractor by an analytic approach,
 we assume that the time average of $\Delta $, denoted by
$\langle \Delta\rangle$, does not change so quickly.
We then discuss only three equations for $\varpi$, $\OmYM$,
and $\Gamma$, giving $\langle \Delta\rangle=\Delta_0$ (constant).
From our numerical analysis, 
we find the following two typical asymptotic behaviors: \\[1em]
~~\,{\bf (i)}\, $\Omega_V\rightarrow$ a finite value ($\lambda<0$)\,,
\\
~~{\bf (ii)} $\Omega_V\rightarrow 0$ ($\lambda>0$)\,.
\\[1em]
$\Gamma$ increases monotonically in our numerical study.
We then do not consider the equation for $\Gamma$ 
to find an approximate asymptotic solution.
For the case (ii), since there is the Hamiltonian constraint 
(\ref{Friedmann2}), $\varpi$ and 
$\OmYM$ are not independent.

We discuss the possible asymptotic solutions separately:

\subsubsection{\rm Case (i)}
For the case (i), the dynamical equations are 
\begin{eqnarray*}
\varpi'&=&{1\over 2}\left(6-\varpi^2\right)
\left(\alpha-\varpi\right)+[2\varpi-3(\alpha+\lambda\Delta_0)]
\OmYM
\\
\OmYM'&=&
\left[-4+\varpi^2+\lambda\varpi\Delta_0+4\OmYM
\right]\OmYM
\,,
\end{eqnarray*}
where the reduced system gives a ``fixed point"
\bea
\varpi={4\over \alpha+\lambda\Delta_0}\,,~~
\OmYM={\alpha(\alpha+\lambda\Delta_0)
-4\over (\alpha+\lambda\Delta_0)^2}
\,.
\ena
Using this ``fixed point", the equation for $\Gamma$ is written as
\bea
\Gamma'={\alpha-\lambda\over \alpha+\lambda\Delta_0}\Gamma ~(>0)
\,.
\ena
This shows a monotonic increase of $\Gamma$, which is confirmed by
 our numerical calculation.

The power exponent of the scale factor and the density parameter 
of the potential are given by
\bea
p={\alpha+\lambda\Delta_0\over 2\alpha}
\,,~~
\Omega_V={3\lambda\Delta_0(\alpha+\lambda\Delta_0)+4
\over 3(\alpha+\lambda\Delta_0)^2}
\,.
\ena

From the positivity 
of density parameters, the following conditions must be imposed:
\bea
\alpha(\alpha+\lambda\Delta_0)\geq 4
\,,~~
3\lambda\Delta_0(\alpha+\lambda\Delta_0)+4\geq 0
\,.
\label{existence_Delta0}
\ena

\begin{figure}[h]
\begin{center}
\includegraphics[scale=.35]{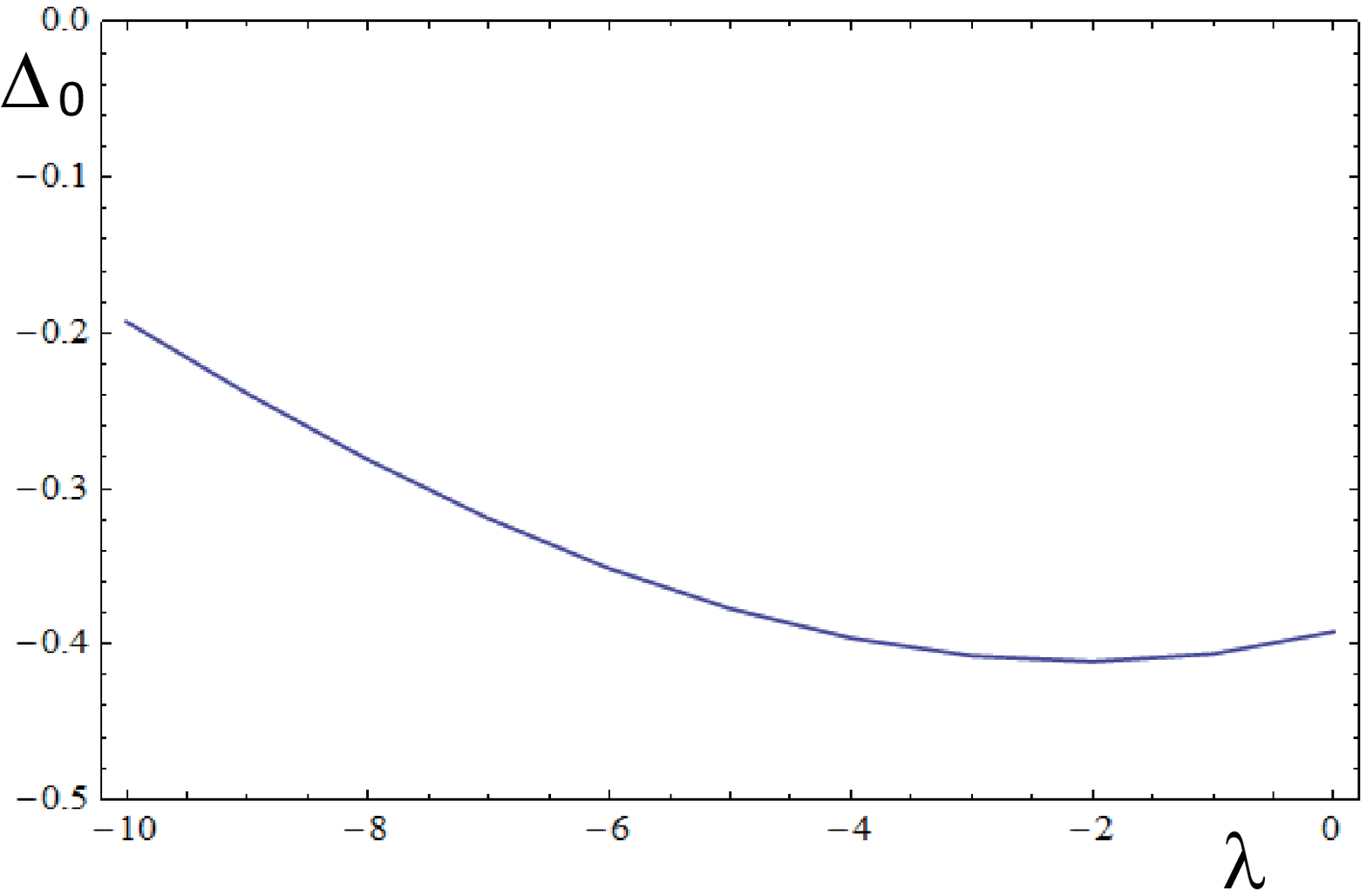}
\\[-1em]
(a)
\\
\includegraphics[scale=.35]{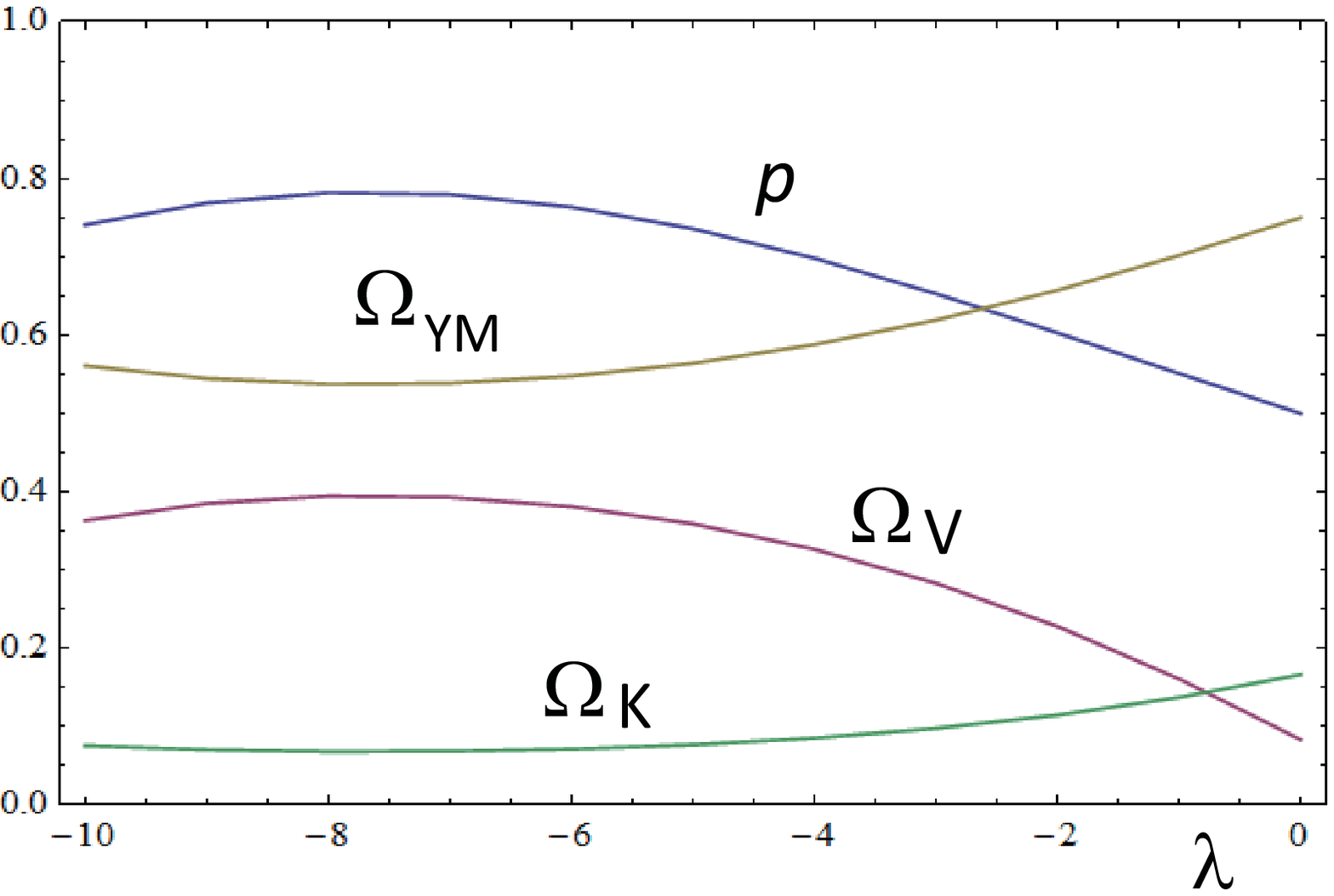}
\\[-1em]
(b)
\caption{The averaged values of $\Delta_0$ (a),
 the power exponent $p$ of the scale factor 
 and density parameters ($\OmYM, \Omega_V$
 and $\Omega_K$) (b)  for $\alpha=4$ and $\lambda<0$. 
}
\label{fig:delta0_YM-}
\end{center}
\end{figure}

Once we know $p$ and $\varpi$ of the background spacetime, 
we can solve the YM equation as shown in Appendix \ref{YM_oscillation}.
Using this solution, we can take an average of $\Delta$.
However the background spacetime depends on $\Delta_0$, which must be 
the same as the above averaged value $\langle\Delta\rangle$.
Hence we need an iterative procedure to find the correct averaged value 
of $\Delta_0$.
In Fig. \ref{fig:delta0_YM-}, we present our result.

As for the stability, we perturb the above two equations, whose
 eigenvalues are given by the two roots of the quadratic equation
\begin{eqnarray*}
&&\omega^2+\left({\alpha+3\lambda\Delta_0 \over 
\alpha+\lambda\Delta_0}\right)\omega
\nonumber \\
&&~~~+{\alpha(\alpha+\lambda\Delta_0)
-4\over (\alpha+\lambda\Delta_0)^2}
\left[
4+3\lambda\Delta_0(\alpha+\lambda\Delta_0)
\right]=0
\,.~~~
\end{eqnarray*}

From the existence condition (\ref{existence_Delta0}),
we find the following stability conditions: for $\alpha>4/\sqrt{3}$,
\bea
-\alpha<\lambda\Delta_0<(\lambda\Delta_0)^{(-)}\,,~~
\lambda\Delta_0>(\lambda\Delta_0)^{(+)}\,,
\ena
where 
\bea
(\lambda\Delta_0)^{(\pm)}:={-\alpha\pm\sqrt{\alpha^2-16/3}\over 2}
\ena
while for $\alpha<4/\sqrt{3}$,
we have only $\lambda\Delta_0>-\alpha$.
The existence condition guarantees that two eigenvalues are negative.
As a result, this ``fixed point" is always  stable, although $\Gamma$ 
diverges monotonically.
We expect the universe in the parameter range of 
 (\ref{existence_Delta0})
 will evolve into this spacetime with the oscillating YM field.
Since $\lambda\Delta_0>0$, we find that $p>1/2$,
which is consistent with our numerical calculations.
Note that for $\lambda=0$, by which we have a scalar field and Yang-Mills 
field without interaction, we find $p=1/2$ as we expect.

These approximate ``fixed point" solutions seem to explain well 
our numerical results.

\subsubsection{\rm Case (ii)}
For the case (ii), using the relation $\OmYM=1-\varpi^2/6$,
we consider the following equation for $\varpi$:

\begin{eqnarray*}
\varpi'&=&-{1\over 6}\left(6-\varpi^2\right)
\left(\varpi+3\lambda\Delta_0\right)
\end{eqnarray*}
The asymptotic solution can be obtained 
as a ``fixed point" in this sytem, which is
\bea
\varpi=-3\lambda\Delta_0
\,.
\ena
It gives 
\bea
\OmYM&=&1-{3\over 2}\lambda^2\Delta_0^2
\,,
\\
p&=&{2\over 4+3\lambda^2\Delta_0^2}~(<1/2)
\,.
\ena

\begin{figure}[h]
\begin{center}
\includegraphics[scale=.35]{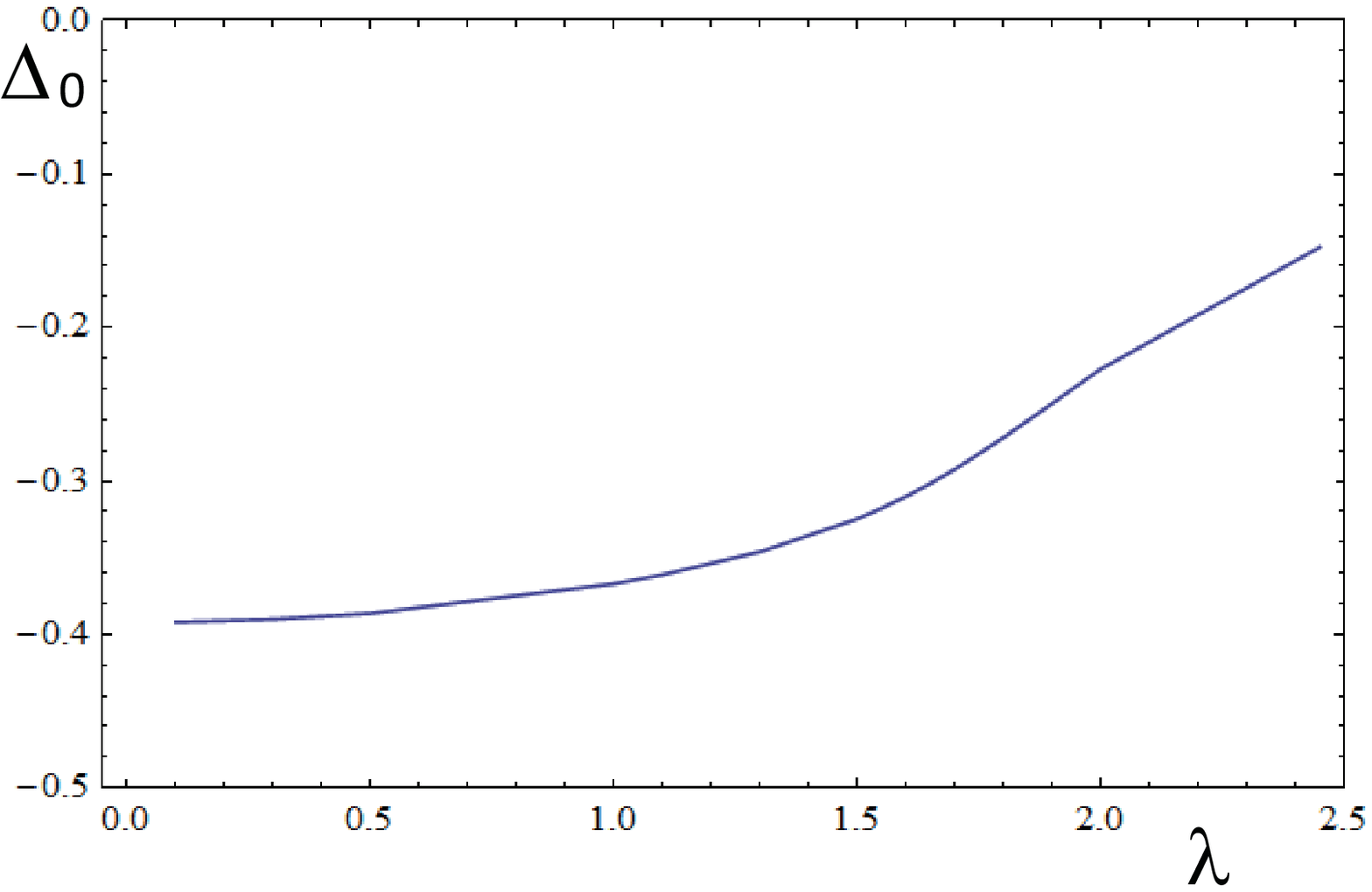}
\\[-1em]
(a)
\\
\includegraphics[scale=.35]{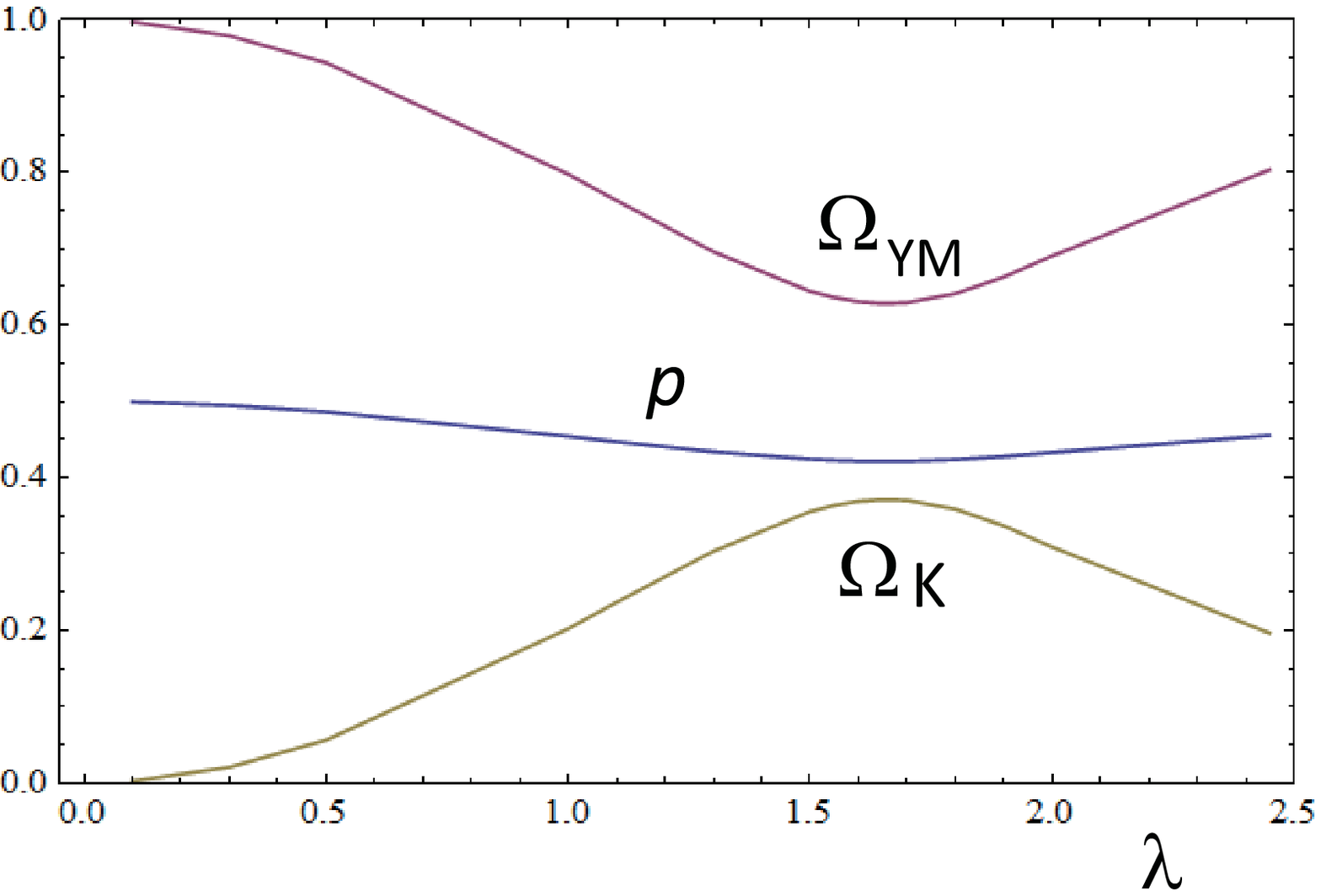}
\\[-1em]
(b)
\caption{The averaged values of $\Delta_0$ (a),
 the power exponent of the scale factor 
$p$ and density parameters ($\OmYM, \Omega_V$
 and $\Omega_K$) (b)  for $\alpha=4$ and $\lambda>0$. 
}
\label{fig:delta0_YM+}
\end{center}
\end{figure}

In this background spacetime, we can also solve the YM equations as given 
in Appendix \ref{YM_oscillation}. 
Using this oscillating solution, we evaluate the averaged value $\Delta_0$.
However, since the background spacetime depends on $\Delta_0$,
we have to find the correct value of $\Delta_0$ iteratively.
In Fig. \ref{fig:delta0_YM+}, we show the result.

Although the qualitative behavior coincides with our numerical result
(for example, $p<1/2$ and $\Gamma$ increases exponentially.), 
it does not reproduce our numerical result quantitatively
For instance, the asymptotic value of $\OmYM$ is $\sim 0.7$ 
in this approximation, but the numerical value is $\sim 0.4$.
A possible source of discrepancy is that the oscillating 
time-scales for $\Delta $ and $\OmYM$ are the same
so that one cannot replace  $\Delta$ by the 
constant averaged value $\Delta_0$ in the 
analysis of the dynamics of $\OmYM$ and $\varpi$,
even though the amplitudes of 
oscillations for those variables is dying away.


\section{Concluding Remarks}
\label{Concluding_Remarks}

We have studied an SU(2) non-Abelian gauge field coupled 
exponentially to a scalar field with an exponential potential,
while making a comparison with the U(1) multiplet case.

We found that the power-law inflation with the magnetic component
of the gauge field ($\BYM$-I) is possible and 
it is an attractor of the present system, if $\lambda>\alpha$
and $\lambda>-\alpha+4/\alpha$.
 The transfer of scalar kinetic energy to the gauge fields through 
the gauge-kinetic coupling makes 
an inflationary solution possible even for a steep potential such as
$\alpha>\sqrt{2}$, which is expected in the unified theories of 
fundamental interactions.

On the other hand, the inflationary solution dominated by the electric 
component ($\EYM$-I) turned out to be unstable in contrast 
to the U(1) multiplet case.
It can be a transient if the initial conditions are tuned.
The attractor of the system is instead the conventional power-law inflation 
($\SYM$-I) if $\alpha <\sqrt{2}$. The YM field with a small amplitude
is oscillating in this background universe.

We have also found new fixed points ($NA_\pm$) in the parameter 
range of $\sqrt{6}<\lambda<\alpha$,
which do not exist in the U(1) multiplet case. 
The fixed point $NA_+$ is an attractor,
while $NA_-$ is unstable.
We have also analyzed the non-inflationary regime,
where the generic feature appears to be the oscillation 
of the YM fields ($O_\pm$).

\begin{figure}[h]
\begin{center}
\includegraphics[scale=.35]{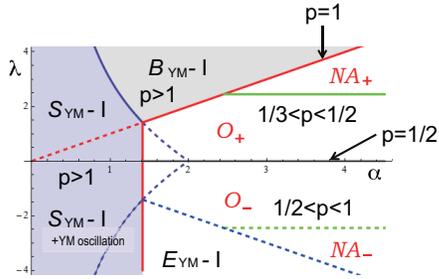}
\caption{The phase diagram in the 
parameter space of the present model. An inflationary phase
is the attractor in the shaded regions. Besides, an attractor solution
of the conventional sense exists for $\lambda >\sqrt{6}$ ($NA_{+}$).
For the rest of the space, the nature of the dynamics is oscillation of
the YM field.}
\label{fig:power}
\end{center}
\end{figure}

We summarize our result in Fig. \ref{fig:power}.

One may wonder whether  those isotropic inflationary solutions are 
stable against anisotropic perturbations.
Since there exist vector fields ($A_\mu^{\rm (a)}$), 
we usually find an anisotropic spacetime just as the case with 
a single U(1) gauge field.
 In order to prove the predictive power of the scenario, we have to show 
that the FLRW universe is obtained as an attractor in anisotropic
Bianchi cosmologies. It is also interesting to know whether 
anisotropic inflation appears in a transient phase and 
its relic is observable or not.
The study of Bianchi universe in the present model is in progress.

Another important subject in the present model is a graceful exit from
a stable inflationary universe, including a reheating mechanism and 
a calculation of density fluctuations. 
In order to leave the power-law inflationary attractor that is
a self-similar scaling solution, within the context of unified theories
of fundamental interactions, we may have the 
following possibilities,
which may also work for the U(1) triplet case:\\
(1) {\bf The moduli fixing}:  After  a certain number of e-folds, 
if we can fix the moduli field $\phi$, the gauge-kinetic coupling 
vanishes. As a result, the inflation with magnetic component ($\BYM$-I) 
will end. \\
(2) {\bf Hybrid-type Inflation}:  If $V_0$ is not
just a constant but depends on another scalar field $\sigma$ as
$V_0={m^2\over 2}\sigma^2$, we find  a dynamics approximated
by the present scenario for the large value of $\sigma$,
 and the end of inflation arrives when $\sigma$ gets small. \\
(3) {\bf Decay of the VEV of YM field}:  The YM field 
may be coupled to other particles.  Through such a coupling, the particles 
can be created quantum mechanically, which will reduce the YM vacuum energy 
($\rhoYM$)\cite{Dimopoulos}. The inflation assisted by YM field will 
eventually end.
 
As for the reheating of the universe, 
for the cases (1) and (2), 
since we have a potential minimum 
around which a scalar field will oscillate, we then  find 
the reheating of the universe.
It is not clear whether we can find the hot Big Bang state via 
the particle production assumed in the case (3).

The density fluctuations have been calculated 
for the case of U(1) triplet, which shows  the leading order 
effect of the background gauge fields is consistent with the 
current observational data\cite{Yamamoto2}. While YM field
is expected to give qualitatively similar results at the linear order, 
there is an interesting prospect of generating non-Gaussianity
through the famous chaotic behaviors that are peculiar to 
the non-Abelian gauge fields\cite{YMchaos,Jin,Murata}.
These subjects are  to be investigated in future works.

\acknowledgments
We would like to thank John Barrow, Gary Gibbons, Keiju Murata, 
Nobuyoshi Ohta, and Paul Townsend for valuable comments.
This work was partially supported by the Grant-in-Aid for Scientific Research
Fund of the JSPS (C)  (No.22540291).
KM would like to thank DAMTP and the Centre for Theoretical Cosmology
for hospitality during this work and Clare Hall for a Visiting Fellowship.
He would also acknowledge a hospitality of APC, where this work was completed.
KY would like to thank the Institute of Theoretical Astrophysics
in the University of Oslo for the support and hospitality.



\newpage

\appendix

\section{Oscillation of the Yang-Mills field in the expanding universe}
\label{YM_oscillation}

In some numerical calculations, we have seen the YM field oscillates 
very rapidly while the background spacetime 
evolves smoothly.
If the energy density of the  YM field is much smaller than 
that of the  scalar field,
the YM field does not contribute to the evolution of the universe.
Even for the case that the YM field energy cannot be ignored,
the oscillation of YM field may not directly affect 
the dynamics of the universe, but its mean value may contribute to 
 the evolution of the universe.
The different time-scales of the YM field oscillation and the evolution
of FLRW universe may allow us to treat
these two separately.
Here we find such an oscillation of the YM field, assuming 
a given background spacetime and evolution of the scalar field. 

Suppose
the background is described by
the following power-law solution:
\bea
a=a_0 t^p\,,~~\phi=\varpi_0 N +\phi_0
\,,
\ena
where $p$ and $\varpi_0$ are constants, and $N=\ln (a/a_0)$
 is the e-folding time.
The equation for the isotropic YM field in this background is given by 
\bea
\ddot A+{p(1+\lambda \varpi_0)\over t}\dot A+{2\gYM^2\over a_0^2}
{A^3\over t^{2p}}=0
\,.
\ena
Changing the variables $t$ and $A$ to $\eta$ and $Z$, which are defined by
\bea
t&=&\left({\eta\over s}\right)^s
\,,
\\
A&=&{a_0\gYM\over \sqrt{2}}
\left({\eta\over s}\right)^{-\lambda p\varpi_0 s/3}
\,Z
\,,
\ena
where
\bea
s={3\over 3-p(3+\lambda \varpi_0)} ,
\label{parameter_s}
\,
\ena
we find the following equation for $Z$:
\begin{widetext}
\bea
{d^2Z\over d\eta^2} 
-{\lambda p\varpi_0\over 9}\left[3(p-1)+2\lambda p\varpi_0
\right]
\left({\eta\over s}\right)^{2[3(p-1)+\lambda p\varpi_0]s/3} Z+Z^3=0
\,.
\label{eq_Z}
\ena
\end{widetext}

Let us discuss the case where $\eta$ increases as $t$ increases, i.e.,
we assume that $s>0$, or equivalently, $3(p-1)+\lambda p\varpi_0<0$.
Hence this term in Eq. (\ref{eq_Z}) 
may drop as $\eta\rightarrow \infty$
($t\rightarrow \infty$). 
Once we ignore the second linear term, we find 
a simple non-linear differential equation
\bea
 {d^2Z\over d\eta^2}+Z^3=0
\,,
\ena
which solves as
\bea
Z=Z_0{\rm cn}\left(Z_0\eta;{1\over \sqrt{2}}\right)
\,,
\ena
where ${\rm cn}(x; k)$ is the Jacobi's elliptic function.
Then the YM field is described in terms of the cosmic time $t$ as
\bea
A={a_0\gYM Z_0\over \sqrt{2}}t^{-\lambda p\varpi_0/3}
{\rm cn}\left(Z_0s \,t^{1/s},{1\over \sqrt{2}}\right)
\,.
\ena
Using this solution, we can evaluate the 
asymptotic behavior of the density parameter and the difference
between magnetic and electric components of the YM field  as
\bea
\OmYM&\propto&t^{2-p(4+\lambda\varpi_0/3)}
\\
\Delta&=&2{\rm cn}^4\left(Z_0\eta;{1\over \sqrt{2}}\right)-1
\,.
\ena
Since the above approximate solution contains the parameter $s$
(\ref{parameter_s}), which depends on the background solution,
there are the following two cases: (1) The background is controlled only by the scalar field.
The YM field is oscillating in the background, but its energy density 
 is too small to affect the evolution of  the universe.
(2) The other case is that the averaged value of $\Delta_0$ as well as
$\OmYM$ give an important contribution onto the background.
In that case, we need an iterative procedure to find the 
correct averaged value $\Delta_0$,
as shown in the main body of the article.

Here we present the averaged value of $\Delta$ and 
the properties of the asymptotic spacetime in the case 
(1). The results for the case (2) are given in \S . \ref{sec:iteration}.
 
\begin{figure}[h]
\begin{center}
\includegraphics[scale=.5]{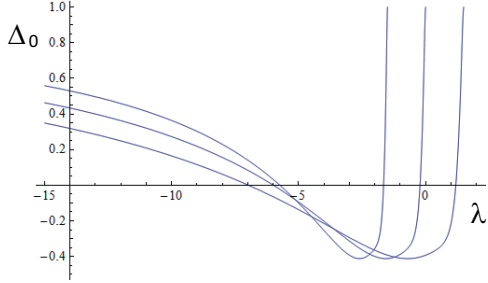}
\caption{The averaged values of $\Delta_0$ for $\alpha=1,
\sqrt{2}$ and $2$. It changes from $-0.4$ to $1$
depending on the value of $\lambda$.
}
\label{fig:delta0_s}
\end{center}
\end{figure}

For inflation driven by a scalar field,
we have $p=2/\alpha^2$ and $\varpi_0=\alpha$.
Then the condition $3(p-1)+\lambda p\varpi_0<0$ is 
\bea
\lambda<{3(\alpha^2-2)\over 2\alpha}
\,.
\ena
This is always satisfied in the range we consider.
The average of $\Delta$ must be taken in terms of the cosmic time $t$.
We show our numerical result in Fig. \ref{fig:delta0_s}.


\begin{thebibliography}{99}
\bibitem{Starobinsky}
A.~A.~Starobinsky,
   Phys.\ Lett.\  {\bf B91}, 99 (1980).

\bibitem{inflation1}
K. Sato, Mon. Not. Roy. Astron. Soc.{\bf 195}, 467 (1981);\\
A. H. Guth, Phys. Rev.  {\bf D23}, 347 (1981).

\bibitem{inflation2}
A. Albrecht and P.J. Steinhardt, Phys. Rev. Lett. {\bf 48}, 1220 (1982);\\
A. D. Linde, Phys. Lett.  {\bf B108}, 389 (1982).

\bibitem{inflation3}
A. D. Linde, Phys. Lett  {\bf B129}, 177 (1983).

\bibitem{inflation4}
See also the following review articles:\\
A. D.~Linde,
  [arXiv:hep-th/0503203v1];
J. Phys.: Conf. Ser. {\bf 24}, 151 (2005)
[arXiv:hep-th/0503195];
Lect. Notes Phys. {\bf 738}, 1 (2008)
 [arXiv:0705.0164 [hep-th]] ;\\
L.~McAllister and E.~Silverstein,
Gen. Rel. Grav. {\bf 40}, 565 (2008)   
[arXiv:0710.2951 [hep-th]];\\
D.~H.~Lyth,
Lect. Notes Phys. {\bf 738}, 81 (2008)
[arXiv: hep-th/0702128].
  
\bibitem{string}
J. Polchinski, ``String Theory",
Cambridge Univ. Press, Cambridge, UK (1998).

\bibitem{Mtheory}
E. Witten, Nucl.\ Phys.\  {\bf B443}, 85 (1995) .

\bibitem{no-go}
G.~W.~Gibbons,
 {\it Proceedings of the GIFT Seminar on Theoretical Physics, San Feliu de
 Guixols, Spain, Jun 4-11, 1984},
 ed. F.~Del~Aguila, {\it et al.} (World Scientific, 1984) pp.~123-146;\\
J.~M.~Maldacena and C.~Nunez,
 Int.\ J.\ Mod.\ Phys.\ A {\bf 16}, 822 (2001)
 [arXiv:hep-th/0007018].

\bibitem{Sbrane1}
P.~K.~Townsend and M.~N.~R.~Wohlfarth,
 Phys.\ Rev.\ Lett.\  {\bf 91}, 061302 (2003)
 [arXiv:hep-th/0303097].

\bibitem{Sbrane2}
N.~Ohta,
 Phys.\ Rev.\ Lett.\  {\bf 91}, 061303 (2003)
 [arXiv:hep-th/0303238];
 Prog.\ Theor.\ Phys.\  {\bf 110}, 269 (2003)
 [arXiv:hep-th/0304172].

\bibitem{Sbrane3}
M.~N.~R.~Wohlfarth,
  Phys.\ Lett.\  {\bf B563}, 1 (2003)
  [arXiv:hep-th/0304089].

\bibitem{Sbrane4}
C.~M.~Chen, D.~V.~Gal'tsov and M.~Gutperle,
  Phys.\ Rev.\  {\bf D66}, 024043 (2002)
  [arXiv:hep-th/0204071];\\
N.~Ohta,
  Phys.\ Lett.\  {\bf B558}, 213 (2003)
  [arXiv:hep-th/0301095].


\bibitem{Brane1}
G.R.~Dvali and S.-H.H.~Tye,
  Phys.\ Lett.\  {\bf B450}, 72 (1999)
  [arXiv;hep-th/9812483];\\
S.B.~Giddings, S.~Kachru and J.~Polchinski,
  Phys.\ Rev.\  {\bf D66}, 106006 (2002)
  [arXiv:hep-th/0105097];\\
S.~Kachru, R.~Kallosh, A.~Linde, and S.P.~Trivedi,
  Phys.\ Rev.\  {\bf D68}, 046005 (2003)
  [arXiv:hep-th/0301240];\\
S.~Kachru, R.~Kallosh, A.~Linde, J.~Maldacena, L.~McAllister and S.P.~Trivedi,
 JCAP {\bf 0310}, 013 (2003),
  [arXiv:hep-th/0308055].

\bibitem{Brane2}
See also the following review article:
S.-H.H.~Tye
 Lect. Notes Phys. {\bf 737}, 949 (2008)
 [arXiv:hep-th/0610221v2].
 

\bibitem{maeda1}
K. Maeda, Phys. Rev.  {\bf D37}, 858 (1988).


\bibitem{HighR1}
H.~Ishihara,
  Phys.\ Lett.\  {\bf B179}, 217 (1986).

\bibitem{HighR2}
K.~Maeda,
   Phys.\ Lett.\  {\bf B166}, 59 (1986);\\
J.~R.~Ellis, N.~Kaloper, K.~A.~Olive and J.~Yokoyama,
   Phys.\ Rev.\  {\bf D59}, 103503 (1999)
   [arXiv:hep-ph/9807482].

\bibitem{HighR3}
K.~Maeda and N.~Ohta,
Phys.\ Lett.\  {\bf B597}, 400 (2004)
[arXiv:hep-th/0405205];
Phys.\ Rev.\   {\bf D71}, 063520 (2005)
[arXiv:hep-th/0411093];\\
K.~Akune, K.~Maeda and N.~Ohta,
Phys.\ Rev.\   {\bf D73}, 103506 (2006)
[arXiv:hep-th/0602242].
\bibitem{HighR4}
K.~Bamba, Z.~K.~Guo and N.~Ohta,
  Prog.\ Theor.\ Phys.\  {\bf 118}, 879 (2007)
  [arXiv:0707.4334 [hep-th]].

\bibitem{HighR5}
K.~Maeda, N.~Ohta, and R.~Wakebe, 
Eur. Phys. J. C {\bf 72}, 1949(2012) 
[arXiv:1111.3251 [hep-th] ].


\bibitem{power-law_inflation}
F. Lucchin and S. Matarrese, Phys. Rev.  {\bf D32} , 1316 (1985).
L.F. Abbott and M.B. Wise, Nucl. Phys.  {\bf B244}, 541 (1987) .

\bibitem{HAL}
J.~J.~Halliwell,
   Phys.\ Lett.\  {\bf B185}, 341 (1987).

\bibitem{Yokoyama_Maeda}
J. Yokoyama, K. Maeda, 
Phys. Lett. {\bf B207} 31 (1988). 


\bibitem{Kitada_Maeda}
Y. Kitada, K. Maeda,
Phys. Rev. {\bf D45} 1416 (1992). 

\bibitem{SG1}
E. Cremmer, S. Ferrara, C. Kounnas and D.V. Nanopoulos,
Phys. Lett. {\bf B133}, 61 (1983) ;
J. Ellis, A.B. Lahanas, D.V. Nanopoulos and K. Tamvakis,
Phys. Len. {\bf B134}, 429 (1984).
\bibitem{Maeda_Nishino}
H. Nishino and E. Sezgin, Phys. Lett. {\bf B144}, 187 (1984);
K. Maeda and H. Nishino, Phys. Lett. {\bf B154}, 358 (1985); 
{\bf B158}, 381 (1985).
\bibitem{SG2}
E. Witten, Phys. Lett. {\bf B155} (1985) 151;
J.P. Derendinger, L.E. Ib$\acute {\rm a}\tilde{\rm n}$ez 
and H.P. Nilles, Nucl. Phys.
{\bf B267}, 365 (1986);
M. Dine, R. Rohm, N. Seiberg and E. Winen, Phys. Lett. {\bf B156}, 55 (1985).

\bibitem{Townsend}
P.K. Townsend, {\it Cosmic Acceleration and M-theory},
in the proceedings of ICMP2003, Lisbon, Portugal (2003)
aiXiv:hep-th/0308149.


\bibitem{Hull-Townsend}

C.M. Hull and P.K. Townsend, Nucl. Phys. {\bf B438}, 109 (1995).

\bibitem{Gibbons-Maeda}

G.W. Gibbons and K. Maeda,
Phys. Rev. Lett. {\bf 104},  131101 (2010).



\bibitem{Kanno1} 
S. Kanno, J. Soda, and M.a. Watanabe, 
J. Cosmol. Astropart. Phys. {\bf 12}, 009 (2009).

\bibitem{Watanabe} 
M.a. Watanabe, S. Kanno, and J. Soda, 
Phys. Rev. Lett. {\bf 102}, 191302 (2010);
Prog. Theor. Phys. {\bf 123}, 1041 (2010).

\bibitem{Kanno2} 
S. Kanno, J. Soda, and M.a. Watanabe, 
J. Cosmol. Astropart. Phys. {\bf 12}, 024 (2010).
 
\bibitem{Yamamoto}
 K. Yamamoto, M.a. Watanabe and J. Soda, Class. Quantum Grav.
 {\bf 29}, 145008 (2012).

\bibitem{Murata} K. Murata and J. Soda, J. Cosmol. Astropart. Phys.
 {\bf 06}, 037 (2011).


\bibitem{Moniz} P. V. Moniz and J. Ward, Classical Quantum Gravity {bf 27}, 
235009 (2010).
 
\bibitem{Do} T.Q. Do, W. F. Kao, and I.C. Lin, Phys. Rev.  {\bf D83}, 
123002 (2011).
 
\bibitem{Hassan} R. Emami, H. Firouzhahi, S. M. Sadegh Movahed, and M. Zerei, 
J. Cosmol. Astropart. Phys. {\bf 02}, 005 (2011).
 
\bibitem{Wagstaff} J. M. Wagstaff and K. Dimopoulos, Phys. Rev.  {\bf D83}, 023523 (2011).
 
\bibitem{Hervik} S. Hervik, D. F. Mota, and M. Thorsurd, J. High Energy Phys.
 {\bf 11}, 146 (2011) .
 

 
\bibitem{Adshead} P. Adshead and M. Wyman, Phys. Rev. Lett. {\bf 108}, 261302
 (2012);
P. Adshead and M. Wyman,  Phys. Rev.  {\bf D86}, 043530 (2012);
E. Martinec, P. Adshead, and M. Wyman, arXiv:1206.2889[hep-th]
 
\bibitem{Anber} M. Anber and L. Sorbo, Phys. Rev.  {\bf D81}, 043534 (2010).
 
 

\bibitem{footnote1}
It is easy to include the curvature term. The Friedmann equation is
\beann
H^2+{k\over a^2}={1\over 3}\left({1\over 2}\dot \phi^2+V+\rhoYM\right)
\,.
\enann
If we have a power-law inflation $a\propto t^p$ ($p>1$),
the curvature term drops faster than the Hubble expansion term as
$k/a^2 \propto t^{-2p}$ and $H^2\propto t^{-2}$.
The curvature term can be ignored asymptotically 
just as the conventional inflationary scenario.
However, it will be important in the non-inflationary 
universe.

\bibitem{Dimopoulos} K. Dimopoulos, G. Lazarides, and J. M. Wagstaff, 
J. Cosmol. Astropart. Phys. {\bf 02}, 018 (2012).

\bibitem{Yamamoto2}
K. Yamamoto, Phys. Rev. {\bf D85}, 123504 (2012).
\bibitem{YMchaos}
G. Z. Baseyna, S. G. Matinyan, G. Z. Savvidy, JETP Lett. {\bf 29}, 587 (1979);
S. G. Matinyan, G. Z. Savvidy, N. G. Ter-Arutyunyan-Savvidi, Sov. Phys. JETP 
{\bf 53}, 421 (1981);
B. V. Chirikov, D. L. Shepelyanskii, JETP Lett. {\bf 34}, 163 (1981);
B. V. Chirikov, D. L. Shepelyanskii, Sov. J. Nucl. Phys. {\bf 36}, 908 (1982);
G. K. Savvidy, Phys. Lett. {\bf B130}, 303-307 (1983);
S. G. Matinyan, E. B. Prokhorenko, G. K. Savvidy, Nucl.
Phys. {\bf B298}, 414 (1988); T. Kawabe, S. Ohta, Phys. Rev. 
{\bf D41}, 1983 (1990); 
T. Kawabe, S. Ohta, Phys. Rev. {\bf D44}, 1274 (1991).
See also the review. 
T. S. Byro, S. G. Matinyan, B. M\"{u}ller, 
{\it 
Chaos and gauge field theory}, (World Scientific, Singapore, 1994).
\bibitem{Jin}
Y. Jin and K. Maeda,Phys. Rev. {\bf D71}, 064007 (2005);
J.D. Barrow, Y. Jin and K. Maeda,
Phys. Rev. {\bf D72}, 103512 (2005).

\end{thebibliography}
\end{document}